\documentclass[preprint,prd,tightenlines,nofootinbib,superscriptaddress]{revtex4-1}

\usepackage{color}
\usepackage{graphicx}
\definecolor{red}{rgb}{1,0,0}
\def\lesssim{\ \hbox{\raise 2pt \hbox{$<$} \kern -13pt
                     \lower 3pt \hbox{$\sim$}}\ }
\def\greatersim{\ \hbox{\raise 2pt \hbox{$>$} \kern -13pt
                     \lower 3pt \hbox{$\sim$}}\ }

\input epsf.tex
\def\desepsf(#1 width #2){\epsfxsize=#2 \epsfbox{#1}}

\def\kt{\ensuremath{k_t}}

\newcommand{\alphasb}{\bar{\alpha}_s}

\newcommand{\Pmax}{p}
\newcommand{\cA}{{\cal A}}

\usepackage{amsmath,bm}
\begin{document}

\hspace*{12.9 cm} {\small DESY 13-261}

\vspace*{1.4 cm} 

\title{Transverse momentum dependent gluon density  \\   
from  DIS precision data}
\author{F.\ Hautmann} 
\affiliation{Dept.\  of Physics and Astronomy, 
University of   Sussex,   Brighton   BN1  9QH}
\affiliation{Rutherford Appleton Laboratory,  Chilton  OX11  0QX}
\affiliation{Dept.\  of  Theoretical Physics, 
University of Oxford,    Oxford OX1 3NP}  
\author{H.\ Jung}
\affiliation{Deutsches Elektronen Synchrotron, D-22603 Hamburg}
\affiliation{Elementaire Deeltjes Fysica, Universiteit Antwerpen, B 2020 Antwerpen}

\begin{abstract}
The    combined measurements  of  proton's  structure functions 
 in deeply inelastic scattering at the  HERA  collider  provide high-precision data 
 capable of constraining  parton density functions  over a wide range of the kinematic variables. 
We perform   fits to these data using transverse momentum 
dependent  QCD factorization and CCFM evolution.  
  The results  of the fits to  precision measurements  are used 
 to make a determination of the nonperturbative 
transverse momentum dependent gluon density function, 
 including experimental and theoretical  uncertainties. 
We  present   an application of this  density function   to vector boson +  jet  production 
processes at the LHC. 
\end{abstract} 


\maketitle

\section{Introduction} 
\label{sec:1}

The  high-precision combined HERA data~\cite{Aaron:2009aa}  
 for   proton's  deeply inelastic  scattering (DIS) 
 structure functions  
constrain   parton density functions (pdfs) 
 over a wide range of the kinematic variables. These  data 
have been used for determinations of the 
collinear pdfs and  related  
studies  at the LHC~\cite{pdf4lhc-alekhin}. 

On the other hand,    QCD  applications to 
multiple-scale scattering problems  and complex final-state 
observables 
require  in general   formulations of 
   factorization~\cite{jcc-book}  which 
involve transverse-momentum dependent  (TMD),  or 
  unintegrated,     
 parton  density and parton decay  
functions~\cite{mert-rog,mulders,jada09,unint09}.   
TMD pdfs  are necessary 
to describe appropriately nonperturbative physics 
  and to control  perturbative  large 
logarithms to higher orders of perturbation 
theory.

The purpose of  this work   is  to 
use  the combined 
DIS data on structure functions~\cite{Aaron:2009aa} and 
charm production~\cite{comb-charm}   for 
determination of   TMD pdfs.  
A general  program  for  TMD pdfs phenomenology 
has been proposed in~\cite{mert-rog}.  Our  work  
has a more limited scope  than this program 
 as we  limit 
ourselves to considering DIS  data in the 
small-$x$ kinematic region.  On the other hand, 
from the point of view of TMD pdfs,   
this region is  interesting   because a  well-defined form
of  TMD factorization holds 
 at high energy~\cite{hef},  which  has  been applied 
 to sum  
small-$x$ logarithmic corrections 
 to DIS to all orders in $\alpha_s$  
at leading  and  
next-to-leading $\ln x$ level~\cite{dis-x-phen,ch94,fl98}. 
Furthermore, given the high precision of  the combined 
data~\cite{Aaron:2009aa,comb-charm}, this analysis provides a 
compelling test of the TMD approach 
and of the  limitations of the 
 logarithmic  approximations used at small $x$.  
This is to be contrasted with earlier 
 analyses~\cite{Jung:2002wn,Hansson:2003xz}   based 
on older and much less precise structure function 
measurements.  

The high-energy  factorization~\cite{hef}  expresses  
the heavy-quark  leptoproduction cross section 
 in terms of the TMD gluon 
density via well-prescribed, calculable perturbative 
coefficients. 
This framework is extended to 
 DIS structure functions  in~\cite{ch94}. 
  Phenomenological applications  of   
this approach   require   matching of
   small-$x$  contributions  
 with  contributions from 
 medium   and large 
$x$~\cite{dis-x-phen,Andersson:2002cf,hj04,hj-ang,ajalt,
Marchesini:1992jw,heraproc92}.  
To do this,  in this work  we  further
develop   the parton branching  
Monte Carlo~\cite{Marchesini:1992jw}  
implementation  of  
 the CCFM  evolution 
equation~\cite{skewang,Catani:1989sg},  which  
 we  include  in the 
 \verb+herafitter+   program~\cite{Aaron:2009aa,herafitter}.  
The  TMD gluon distribution at  the initial scale 
$q_0$  of  the evolution 
 is  determined from fits to  DIS data, including charm production. 

We perform  fits to  measurements of the
$F_2$  structure function~\cite{Aaron:2009aa}   
in the range  $x<0.005$, $Q^2>5$~GeV$^2$ and 
to measurements of  the charm  structure function 
$F_2^{({\rm{charm}})} $~\cite{comb-charm} in the range 
$Q^2 > 2.5$~GeV$^2$. We obtain 
good fits to $F_2$ and $F_2^{({\rm{charm}})} $, and 
   we      make   a determination of  
the TMD gluon density (as well as   of the charm  mass $ m_c$ 
  and of $\Lambda_{\rm{QCD}}$, or 
 the strong coupling $\alpha_s$)  based on these. 
We find that the best fit to 
$F_2^{({\rm{charm}})} $ 
gives 
$\chi^2$ per degree of freedom 
$\chi^2/ndf \simeq 0.63$, and the best fit to 
$F_2$  gives $\chi^2/ndf \simeq 1.18$. 
The method allows one to  assign 
 experimental and theoretical uncertainties to   
pdfs.  We give  results for  these   
different kinds of  pdf uncertainties. 
We also carry out an application of the 
 TMD gluon density 
determined  from HERA data  fits to  LHC physics,  
 by computing  predictions for  $W$-boson  +   jet  
production  in proton-proton collisions. 

The paper is organized as follows.     In Sec.~\ref{sec:2}  we  
summarize the main elements of the approach based on 
high-energy factorization and evolution, and discuss  
a few   details on its   implementation. 
 In Sec.~\ref{sec:3} we describe the fits to charm and 
DIS precision data. 
We discuss $F_2$, $F_2^{({\rm{charm}})} $,  
 the TMD gluon density determination 
 and associated  uncertainties.  
In Sec.~\ref{sec:4}  we illustrate 
the use of TMD gluon density at the  LHC.  We give 
conclusions in   Sec.~\ref{sec:5}.

\section{Factorization and  evolution} 
\label{sec:2}

In the  framework of high-energy factorization~\cite{hef}  
  the deeply inelastic   
scattering 
cross section is written as a  convolution in 
both longitudinal and transverse momenta   of 
the TMD  parton density function 
${\cal  A}\left(x,\kt,\mu\right)$    
 with   off-shell partonic 
matrix elements, as follows 
\begin{equation}
 \sigma_j  ( x , Q^2 )  = \int_x^1  
d z  \int d^2k_t \ 
\hat{\sigma}_j( x   ,  Q^2 ,  {    z}   ,  k_t ) \ 
 {\cal  A}\left( { z} ,\kt,  \mu \right)  .  
\label{kt-factorisation}
\end{equation}
Here 
$x$ and $Q^2$ denote the Bjorken variable and photon 
virtuality, and the DIS cross sections 
$\sigma_j$,  with $j= 2 , L$, are related to the customary 
DIS structure functions $F_2$ and $F_L$ 
 by~\cite{eswbook}    
\begin{equation}
\sigma_2 = {{  4 \pi^2 \alpha } \over  Q^2 } \   F_2  \;\; , \;\;\;\;\;  
   \sigma_L = {{  4 \pi^2 \alpha } \over  Q^2 } \   F_L  \;\; ,  
\label{customary2L}
\end{equation}  
where $\alpha$ is the electromagnetic coupling. 
The factorization formula (\ref{kt-factorisation})  
allows one to  resum  
logarithmically enhanced $ x \to 0 $ contributions  
  to all orders in perturbation theory,  
both in the  hard 
scattering coefficients and 
in  the parton evolution,  taking fully into account the 
dependence on the factorization scale $\mu$ and on the 
factorization scheme~\cite{ch94}.  Explicit  
evaluations have been carried out  
through next-to-leading logarithmic 
accuracy~\cite{dis-x-phen,ch94,fl98} 
in  DIS  at $ x \to 0$.  

The physical origin of the logarithmically enhanced $ x \to 0 $ 
corrections   at   higher loops lies  in  the contribution from  regions 
 not ordered in  initial-state 
 transverse momenta to  the QCD  multi-parton matrix elements. 

The perturbative  higher-loop   corrections 
are generally found 
to be  large at   small  $x$. Consider for instance  the   gluonic 
hard-scattering coefficient function  $C_2^g  
( x , \alpha_s, Q^2 / \mu^2 ) $ for the DIS structure 
function $F_2$~\cite{eswbook}.   
 Taking Mellin moments with respect to $x$, 
\begin{equation}
\label{c2-mom-def}
  C_{ 2 ,  N}^g (\alpha_s , Q^2/ \mu^2 ) 
=  \int_0^1 dx \ x^{N-1}   \   C_2^g  ( x , \alpha_s, Q^2 / \mu^2 )  \; , 
\end{equation}
the perturbative expansion of    $  C_2^g 
$  for  $ N \to 0 $, resummed to all orders in 
$\alpha_s   $ via       Eq.~(\ref{kt-factorisation}), 
is given 
at scale $\mu^2 = Q^2$   in the   $ {\overline{\rm MS}} $ minimal subtraction  scheme 
 by~\cite{ch94} 
\begin{eqnarray} 
\label{c2-mom-res} 
 &&   C_{ 2 ,  N}^g (\alpha_s , Q^2/ \mu^2 = 1  ) 
\nonumber\\ 
&& 
 =  { \alpha_s \over { 2 \pi}  }  T_R N_f {2 \over 3} 
\left\{  1 + 1.49 \  {  {\overline \alpha}_s \over N }  
+ 9.71 \left(  {  {\overline \alpha}_s \over N } \right)^2  
+ 16.43  \left(  {  {\overline \alpha}_s \over N } \right)^3 + 
{ \cal O }      \left(  {   \alpha_s \over N } \right)^4  
\right\}  \;  , 
\end{eqnarray}
where $  {\overline \alpha}_s  =  \alpha_s  C_A / \pi$, 
$C_A = 3$, $T_R = 1 / 2$. 
 The $N \to 0$ poles $ \alpha_s 
  (     \alpha_s / N  )^k $, $ k \geq 1$,  
correspond in $x$-space 
via Eq.~(\ref{c2-mom-def})  to next-to-leading-logarithmic  
higher-loop corrections $ \alpha_s^2   
 (\alpha_s  \ln  x )^{k-1}  $.  
The first two terms in 
Eq.~(\ref{c2-mom-res}) are the 
leading-order (LO)~\cite{cfp}   
and next-to-leading-order (NLO)~\cite{vannee} 
small-$x$  contributions to $C_2$. 
The next two terms are the 
three-loop and four-loop small-$x$ contributions. 
The three-loop coefficient   agrees with  the  complete  
next-to-next-to-leading-order (NNLO) 
calculation~\cite{mochetal}.  
The   three-loop and four-loop terms   are 
logarithmically enhanced compared to lower orders. 
 Moreover,  their 
 numerical  coefficients are   significantly 
   larger than the one-loop and two-loop ones. 
Analogous results were obtained in~\cite{ch94}   for 
the coefficient function  $C_L$  of the  longitudinal 
structure function, and confirmed through three loops by the 
NNLO  
calculation~\cite{mochetal-L}.

 Given these results, there is little 
theoretical justification for  treating small-$x$ DIS by 
truncating    the perturbative 
 expansion to  fixed NLO (or NNLO) 
 level.  Thus, although phenomenologically 
 successful in giving very good fits to 
 structure function data,  
  fixed-order  perturbative approaches  are  
 theoretically disfavored, and   cannot   be 
 expected to  describe the physics of the scaling 
  violation in the region of  low values of $x$, where 
 transverse-momentum  ordering does not apply. 
 
 The motivation of  this work is to take a  
 quantitative step toward going beyond fixed-order 
 phenomenology,  and confront our results with the high-precision 
 combined data~\cite{Aaron:2009aa,comb-charm}. 
The approach of this work is based on  
\begin{itemize}
\item including  the 
hard-scattering kernels ${\hat \sigma}_j$ of 
 Eq.~(\ref{kt-factorisation}),    
whose $k_t$-dependence, once convoluted 
with  the  gluon unintegrated Green's  
function~\cite{hef,ch94},  controls the 
all-order resummation of the higher-loop 
terms $ \alpha_s^2   
 (\alpha_s  \ln  x )^{k-1}  $ in the structure 
 functions $F_2$ and $F_L$;  
 \item including  evolution  of the 
transverse momentum dependent gluon density 
${\cal  A} $ by 
combining    the 
    resummation of  small-$x$ logarithmic 
    contributions~\cite{lipatov}  
 with   medium-$x$  and large-$x$ 
 contributions to parton 
splitting~\cite{dglaprefs}.   
\end{itemize}
This     is done via   a 
parton branching  Monte Carlo 
implementation   
of  the CCFM  evolution equation~\cite{skewang,Catani:1989sg}, 
which we  develop based on~\cite{Marchesini:1992jw},  
  and  make available  within the  
\verb+herafitter+   program~\cite{Aaron:2009aa,herafitter}.  

In the remainder of this section we briefly 
 recall the basic  elements 
of this approach and give a few technical details.  
We start  in Subsec.~\ref{subsec:2a} 
 by  recalling the main features of 
  TMD matrix elements and evolution. In 
Subsec.~\ref{subsec:2b} we include the 
unintegrated valence quark density according to the 
method~\cite{Deak:2010gk}. In Subsec.~\ref{subsec:2c}  we 
discuss  aspects 
 of the numerical implementation. 
We give  comments on the general  approach  
in Subsec.~\ref{subsec:2-comment}.

\subsection{TMD matrix elements and evolution} 
\label{subsec:2a} 

The DIS transverse-momentum dependent partonic 
 cross sections  ${\hat \sigma}_j$   
 are evaluated  in the second paper of~\cite{hef} for $ j = 2 $ 
 and in~\cite{heraproc92} for  $ j = L $, including the effects of finite 
 quark masses.   The small-$x$  resummation for  
  DIS structure functions    based on these results 
is carried out  in~\cite{ch94}. 
\begin{figure}[htb]
\begin{center} 
\includegraphics[scale=0.4]{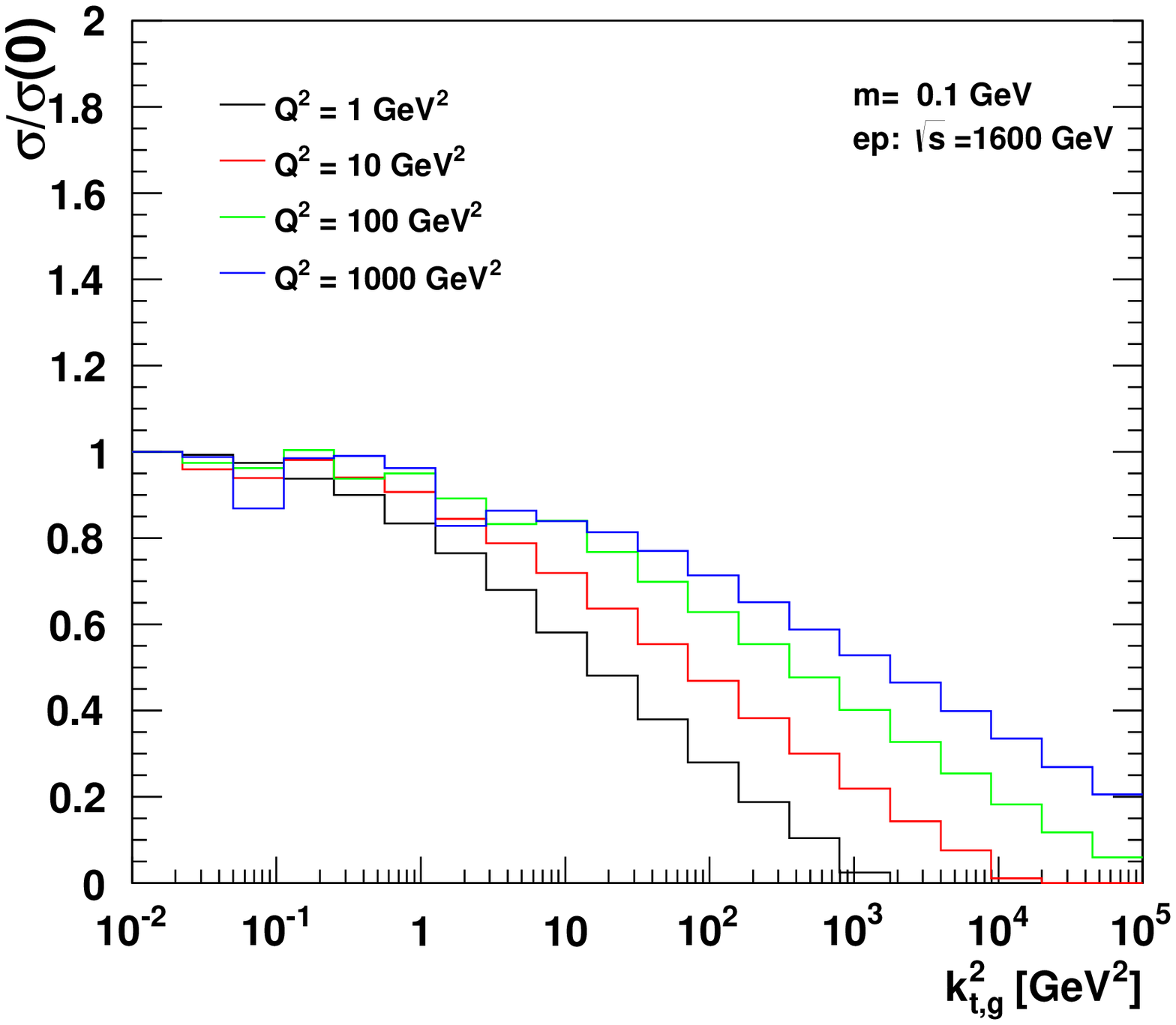}
\includegraphics[scale=0.4]{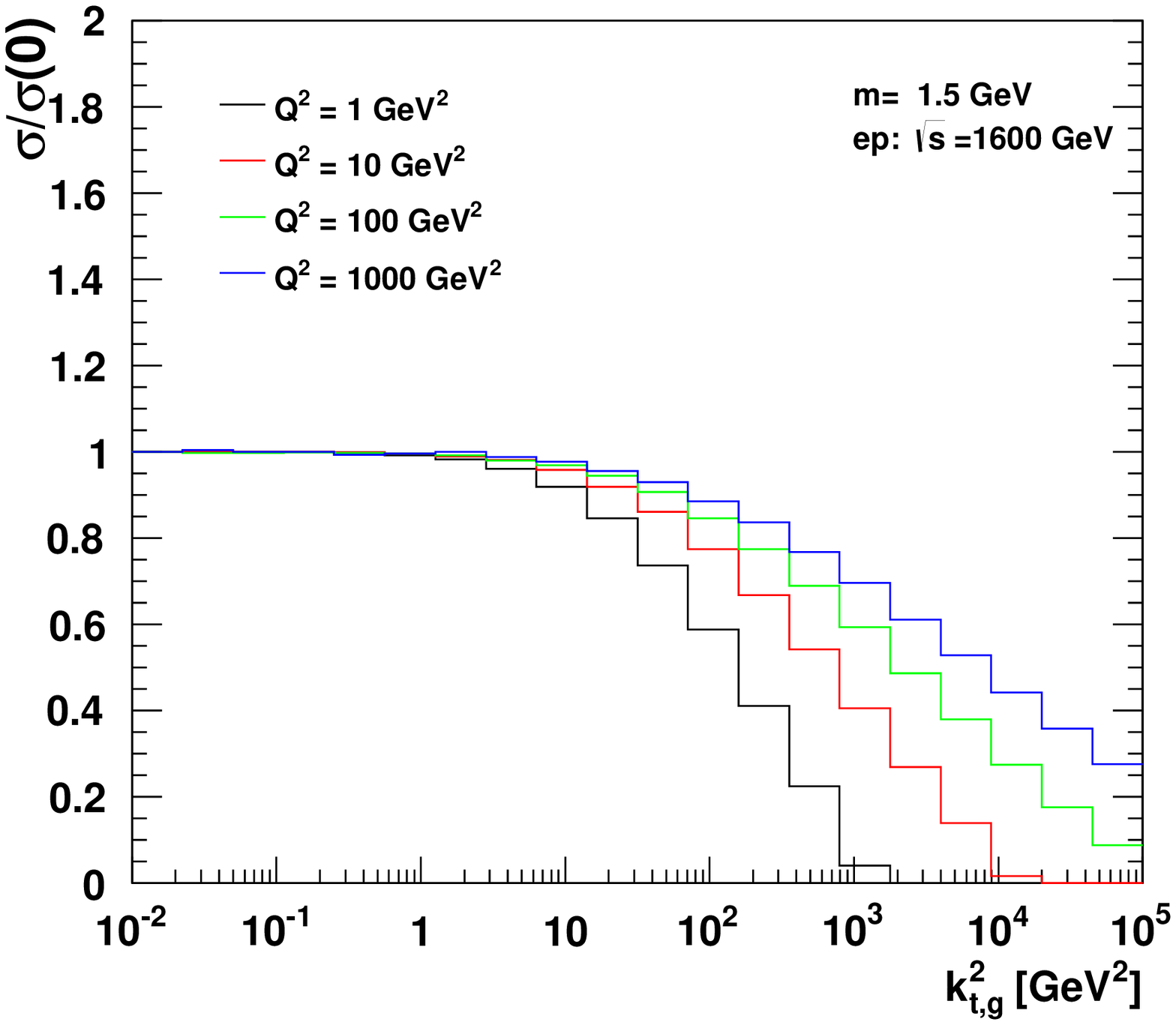}
\includegraphics[scale=0.4]{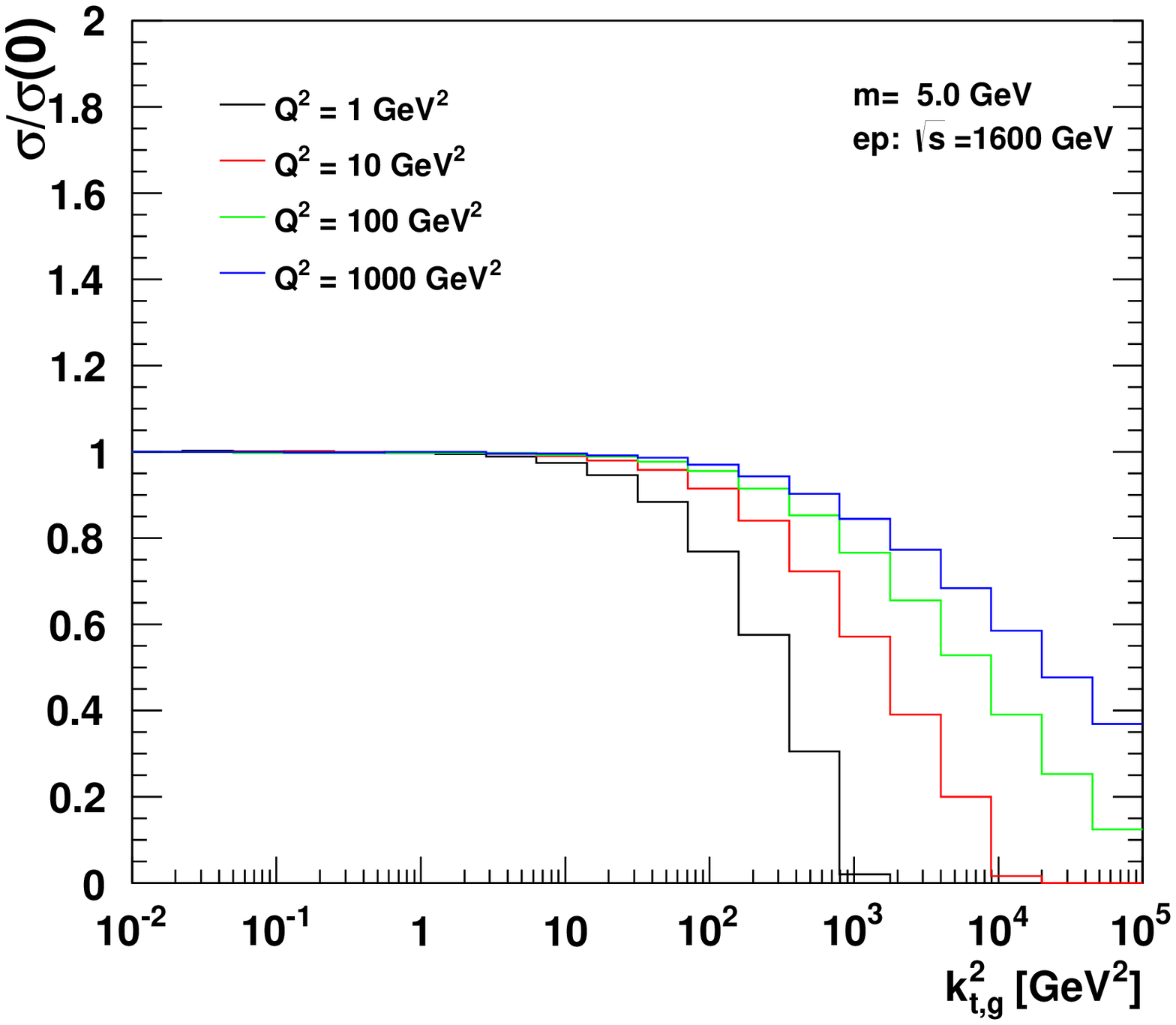}
\caption{\it  Transverse momentum dependence of the  
partonic kernel ${\sigma}_2 $ for the off-shell process 
$ \gamma ( q ) + g ( k ) \to 
q ( p ) + {\bar q } ( p^\prime) $  
at different values of $Q^2$ and quark masses. 
We set $x = 10^{-2}$.}
\label{Fig:ktmatrel}
\end{center}
\end{figure}
Let us  describe the off-shell 
process $ \gamma ( q ) + g ( k ) \to 
q ( p ) + {\bar q } ( p^\prime) $ 
in terms of lightcone momenta 
\begin{equation} 
\label{p1-and-p2} 
p_1^\mu = \sqrt{S/2} \ ( 1 , 0, 0_t)    \;\; , 
\; \;\;\;  p_2^\mu = \sqrt{S/2} \ ( 0 , 1, 0_t)  
  \;\; ,  \; \;\;\;  2 p_1 \cdot p_2 = S \;\; , 
\end{equation} 
where, for any four-momentum, 
 $ p^\mu = ( p^+ , p^- , p_t ) $, with 
 $ p^\pm =  ( p^0 \pm p^3 ) / \sqrt{2}$.  
In the high-energy limit  we have 
\begin{equation} 
\label{q-and-k} 
q^\mu= y p_1^\mu  + \bar{y} p_2^\mu+ q^\mu_t  \;\; , 
\; \;\;\;  k^\mu \simeq  z p_2^\mu + k_t^\mu  \;\;  . 
\end{equation} 
In terms  of   lightcone and transverse    
momentum components, we have $ Q^2 = - q_t^2/(1-y)$, 
$x =Q^2 / ( y S )$. 

In Fig.~\ref{Fig:ktmatrel} we  plot  the  
partonic kernel $ {\sigma}_2 $, normalized to the 
value for $k_t = 0$,  as a function of  
$ k_{t , g}^2 = - k_t^2$ 
 at fixed $x$ and $Q^2$,   
for various values of quark masses.  
As   shown in~\cite{hef,heraproc92},    
the large-$k_t$ tail of the kernels in 
Fig.~\ref{Fig:ktmatrel}, once it is folded with 
unintegrated gluon 
distributions~\cite{skewang,lipatov,jaro}, is 
responsible for 
the logarithmically enhanced higher-loop contributions 
at high energy. 

For evolution of the  TMD gluon density, we require 
that in the limit $x \to 0 $ this evolves with the full 
BFKL anomalous dimension~\cite{lipatov}.  The   
CCFM evolution 
equation~\cite{skewang,Catani:1989sg}  is an exclusive  equation 
which satisfies  
this property (see e.g.~appendix C of~\cite{skewang},  
section~7 of the first paper in~\cite{Catani:1989sg}, 
section~III of~\cite{kwieperla}) 
and,  in addition,   includes  finite-$x $ 
contributions to parton splitting.  It  incorporates 
soft gluon coherence for any value of $x$. 
  The evolution equation 
for the TMD gluon density $\cA(x,\kt,\Pmax)$, 
depending on 
$x$, $\kt$ and the evolution variable $\Pmax$, 
reads 
\begin{eqnarray}
\label{uglurepr1}
  {\cal A} ( x , \kt , p  ) & = & 
  {\cal A}_0 ( x , \kt , p  ) + 
\int { {dz} \over z} \int { { d q^2} \over q^2} \ 
\Theta   (p - z  q)  
\nonumber\\
& \times & 
 \Delta    (p , z  q) 
\ {\cal P} ( z, q, \kt)   
\   {\cal A} 
 ( { x \over z} , \kt  + (1-z) q, q )
 \hspace*{0.3 cm} .        
\end{eqnarray} 
The first term in the right hand side of Eq.~(\ref{uglurepr1}) 
is the contribution of the 
non-resolvable branchings between  the starting scale 
$q_0$ and  the evolution scale $p$, 
 and is given by 
\begin{equation}
\label{uglurepr2}
  {\cal A}_0 ( x , \kt , p  ) =  {\cal A}_0 ( x , \kt , q_0 ) 
 \ \Delta (p , q_0) 
 \hspace*{0.3 cm}    ,   
\end{equation} 
where $\Delta$ is the Sudakov form factor, and 
 ${\cal A}_0 ( x , \kt , q_0 )$     is the 
 starting  distribution   
at scale $q_0$. 
The integral term in the right hand side of Eq.~(\ref{uglurepr1}) 
gives the \kt-dependent branchings in terms of the 
 Sudakov form factor $\Delta$ and unintegrated 
  splitting function ${\cal P}$.  The  gluons' average momentum does not change 
with $p$.

The CCFM evolution equation can be written in a differential form,
 best suited for the backward evolution approach adopted in the Monte
Carlo generator~\cite{Jung:2000hk,Jung:2010si}, as 
\begin{equation}
\Pmax^2\frac{d\; }{d \Pmax^2} 
   \frac{x \cA(x,\kt,\Pmax)}{\Delta_s(\Pmax,q_0)}=
   \int dz  \   \frac{d\phi}{2\pi}\,
   \frac{\tilde{P} (z,\Pmax/z,\kt)}{\Delta_s(\Pmax,q_0)}\,
 x'\cA(x',\kt',\Pmax/z) ,
\label{CCFM_differential}
\end{equation} 
where the splitting variable  $x'$  is given by 
$x'=x/z$,  ${\kt}' = q_t (1-z)/z + {\kt}$,  and $\phi$ is the 
 azimuthal angle  of $q_t$.
The  Sudakov form factor $\Delta_s$ is given by  
\begin{equation}
\Delta_s(\Pmax,q_0) =\exp{\left(
 - \int_{q_0^2} ^{\Pmax^2}
 \frac{d q^{2}}{q^{2}} 
 \int_0^{1-q_0/q} dz \ \frac{\alphasb(q^2(1-z)^2)}{1-z}
  \right)} ,
  \label{Sudakov-delta}
\end{equation}
with ${\overline \alpha}_s=  C_A \alpha_s / \pi =3 \alpha_s  / \pi$. 

The  splitting function $\tilde{P}_g (z_i,q_i,k_{ti})$ for branching $i$ 
is given by~\cite{Hansson:2003xz} 
\begin{eqnarray}
\tilde{P}_g (z_i,q_i,k_{ti})
& = & \alphasb(q^2_{i}(1-z_i)^2)  \  \left( 
 { 1 \over  {1-z_i}}  - 1   + { { z_i (1 - z_i)  } \over 2 }   \right)  
\nonumber\\ 
&  +  & 
\alphasb(k^2_{ti})   
 \ \left(  { 1 \over z_i}   - 1   + { { z_i (1 - z_i)  } \over 2 } 
\right)   \   \Delta_{ns}(z_i,q^2_{i},k^2_{ti})  
\label{Pgg}
\end{eqnarray}
where  $\Delta_{ns}$ is   
the non-Sudakov form factor    
defined by 
\begin{equation}
\log\Delta_{ns} =  -\alphasb(k^2_{ti})
                  \int_0^1 \frac{dz'}{z'} 
                        \int \frac{d q^2}{q^2} 
              \Theta(k_{ti}-q)\Theta(q-z'q_{ti}).
                  \label{non_sudakov}                   
\end{equation}

Quark masses  are treated 
in the fixed flavor number scheme. 
We  include the   two-loop running coupling 
$\alpha_s$, and we apply the kinematic 
 consistency constraint~\cite{Kwiecinski:1996td,Andersson:1995ju} in  the  $g\to gg$ splitting function~\cite{Andersson:2002cf}, 
 given by~\cite{Kwiecinski:1996td}  
\begin{equation}
q_t^2   <   \frac{(1-z) \kt^2}{ z}  . 
\label{consistency-constraint}
\end{equation}

The evolution  (\ref{uglurepr1})  
  of the TMD density implies that 
regions of transverse momenta below  
 $q_0$ can be reached. 
In this region  the branching  is performed  
purely by the Sudakov  form factor.

\subsection{Unintegrated valence quark density} 
\label{subsec:2b} 

Previous determinations of parton distributions based on  the
CCFM evolution have included only the gluon 
density~\cite{Jung:2002wn,Hansson:2003xz}.  
In this work we include  valence quarks using the 
method of~\cite{Deak:2010gk}. 
We  consider the branching evolution equation   at the 
 transverse-momentum dependent level  
 according to 
\begin{eqnarray}
x{ Q_v} (x,k_t,p ) &=&  x{ Q_v}_0 (x,k_t,p ) + \int \frac{dz }{z} 
\int \frac{d q^2}{ q^{2}} \Theta(p - zq) 
\nonumber\\ 
& \times&  \Delta_s  (  p ,  zq)    
P  (z,  q , k_t) \  x{Q_v}\left(\frac{x}{z},k_t +  (1-z) q , q \right)     \;\; , 
\label{integral} 
\end{eqnarray}  
where   $p$ is  the evolution  scale.  The quark splitting function  $P$ is  
given by 
\begin{eqnarray}
P   (z,  q ,  k_t)   &=& {\bar \alpha_s} \left(q^2   (1-z)^2 \right) 
 \frac{1+z^2}{1-z}    \;\;  , 
\label{splitt}
\end{eqnarray}
with $\alphasb=C_F \alpha_s  /  \pi $.  
In Eqs.~(\ref{integral}),(\ref{splitt})    
 the non-Sudakov form factor is not included,  
unlike the CCFM kernel  given in the appendix~B of the first  paper in~\cite{Catani:1989sg},
because  we only associate this 
  factor  to   $1/z$ terms. 
  The term $x{ Q_v}_0$ in 
    Eq.~(\ref{integral})  is the contribution of the non-resolvable branchings 
between  starting scale $q_0$ and evolution scale $p$, given by 
\begin{equation}
 x{ Q_v}_0 (x,k_t,p )   =    x{ Q_v}_0 (x,k_t,q_0 )  \Delta_s  (  p ,  q_0)   \;\; ,   
  \label{Q0term}
\end{equation}
where $ \Delta_s$ is the Sudakov form factor, and the starting distributions  at scale $q_0$  are 
parameterized,   using  the CTEQ 6.6~\cite{nado08} 
    $u$ and $d$ valence quark 
distributions,  as 
\begin{equation}
    x{ Q_v}_0 (x,k_t,q_0 ) =    x{ Q_v}_{\rm{CTEQ66pdf}} (x,q_0 ) \ 
    \exp[ - k_t^2 / \sigma^2 ]   \;\;   .    
\label{gauss}
\end{equation}
In the numerical calculations that follow 
we will take  $ \sigma^2  =  q_0^2 / 2 $.  
For every $p$  we ensure that the flavor sum rule is fulfilled. 

\subsection{Numerical implementation} 
\label{subsec:2c}

 CCFM evolution cannot easily be  written in an analytic closed form. For this 
reason  a Monte Carlo method is employed, based 
on~\cite{Marchesini:1992jw}. 
The Monte Carlo solution is  however time-consuming, and 
cannot be used in a straightforward manner  in a fit program. 
Here we proceed as follows. 
First a kernel $ {\cal \tilde A}\left(x'',\kt,\Pmax\right) $ is determined from the Monte Carlo  solution of the CCFM evolution equation, and then 
this  is folded with the non-perturbative starting 
distribution ${\cal A}_0 (x)$, following the  convolution  method  
 introduced   in~\cite{hj1206}:  
\begin{eqnarray}
x {\cal A}(x,\kt,\Pmax) &= &x\int dx' \int dx'' {\cal A}_0 (x') {\cal \tilde A}\left(x'',\kt,\Pmax\right) 
 \delta(x' 
x'' - x) 
\nonumber  
\\
& = & \int dx' {{\cal A}_0 (x') }  
\cdot \frac{x}{x'} \ { {\cal \tilde A}\left(\frac{x}{x'},\kt,\Pmax\right) } 
\end{eqnarray}
The kernel  ${\cal \tilde A}$ incorporates all of 
the dynamics of the evolution, including   Sudakov form factors and splitting functions.  It 
 is determined on a grid of $50\otimes50\otimes50$ bins 
in $ x,  \kt,  \Pmax$.   The binning in the  grid is 
logarithmic,  except for  the longitudinal variable 
  $x$ where  we use 40 bins in logarithmic 
spacing below 0.1, and 10 bins in linear spacing above 0.1. 

The calculation of the cross section according to Eq.~(\ref{kt-factorisation}) involves a multidimensional Monte Carlo integration which is time consuming and suffers from numerical fluctuations.  This cannot be employed directly in a fit procedure involving the calculation of numerical derivatives in the search for the minimum. Instead 
 the following procedure is applied:
\begin{eqnarray}
\sigma(x,Q^2) & = & \int_x^1 d x_g {\cal A}(x_g,\kt,\Pmax) \hat{ \sigma}(x,x_g,Q^2) 
\nonumber\\
 & = & \int d x_g\; dx'\; dx'' {\cal A}_0 (x') {\cal \tilde A}(x'',\kt,\Pmax)\cdot \hat{ \sigma}(x,x_g,Q^2) \cdot \delta(x' \,x'' -x_g) \nonumber\\
 & = & \int_x^1 dx' {\cal A}_0 (x') \cdot \int_{x/x'}^1 dx''  {\cal \tilde A}(x'',\kt,\Pmax) \cdot \hat{ \sigma}(x,x'\,x'',Q^2) \nonumber\\
  & = & \int_x^1 dx' {\cal A}_0 (x') \cdot \tilde{ \sigma}(x/x',Q^2) \label{final-convolution}
 \end{eqnarray}
Here, first $ \tilde{ \sigma}(x',Q^2)$ is calculated numerically with a Monte Carlo integration on a grid in $x$ for the values of $Q^2$ used in the fit. Then the last step in Eq.(\ref{final-convolution})  is performed with a fast numerical gauss integration, which can be used in standard fit procedures.

\subsection{Comparison with  other  approaches and outlook} 
\label{subsec:2-comment}

The  approach described above, which we are going to confront  in the 
next section with  high-precision DIS measurements,  
can be compared with other approaches to  DIS data  
  in the literature. 
On one hand, it can be contrasted with   descriptions 
of data  based on the 
DGLAP equation~\cite{dglaprefs}  
 at fixed perturbative order, e.g.~NLO or NNLO (see for 
 instance~\cite{pdf4lhc-alekhin} and references therein).  
As recalled at the beginning of this section, these  descriptions, 
however successful  phenomenologically, 
 have little theoretical justification at small $x$, due to the 
structure of the perturbative expansion for $x \to 0$. 
The need to go beyond fixed-order truncations of 
perturbation theory   leads us  to  employ  
a TMD  formalism.   In particular, 
in the  approach of this paper   
the transverse momentum dependence of the gluon 
density  arises both from 
perturbative and from nonperturbative processes. 
Both the kernel and the initial condition 
of  the evolution equation   are  $k_t$-dependent. 
These different sources of  $k_t$-dependence have distinct 
physical effects, for instance  on  the angular distributions of 
associated jet final states in DIS, as 
analyzed in detail in~\cite{hj-ang}. 
This  feature    
 can be contrasted with  TMD approaches    
   focusing  on  nonperturbative 
 $k_t$-dependence, see e.g.~\cite{mulders}.

On the other hand, the 
 approach  of this paper can be 
compared  with  approaches based on the  
BFKL equation~\cite{lipatov}.   
Recent fits to DIS data have   been performed 
 in this context~\cite{sabioetal,henrietal,genyaetal}. 
Compared to these works, 
the main theoretical inputs  
in the  present paper 
are 
\begin{itemize}
\item the use of TMD matrix elements  
which can be directly related with the resummation 
of DIS coefficient functions as in Eq.~(\ref{c2-mom-res}), and 
\item the use of the CCFM evolution  for the 
TMD parton density rather than the BFKL evolution.   This  
 results in  distinctive properties  due to 
soft gluon coherence    of the  final states 
contributing to DIS~\cite{Andersson:2002cf}. 
\end{itemize}
A further,   distinctive  feature  of  the   framework   employed in 
  this paper is that 
the  gluon distribution obtained from DIS fits can be directly used 
to make predictions for final states at the LHC, as 
 we do for example  in Sec.~\ref{sec:4}  for 
$W$-boson    production associated with jets.

Our   approach relies on  
 perturbative factorization theorems, which 
classify higher-order  corrections 
 according to the logarithmic hierarchy  
based on  high  $Q^2$ and low  $x$. 
For this reason  in the next section we will apply this 
 approach 
 to $F_2$ structure function   measurements 
in a range $Q^2 > {\overline Q}^2 $, $ x < {\overline x}$,  
where we choose $ {\overline Q}^2  = 5 $ GeV$^2$, 
$ {\overline x} = 5 \cdot 10^{-3}$. 
For asymptotically small $x$ one expects 
the operator product expansion 
 to  break down and 
 DIS to   become dominated by 
nonperturbative physics.   
 Thus the   low-$Q^2$ region could 
 require methods beyond the ones applied in 
this work,  see e.g.~\cite{fh06,hs07}. 
On the other hand,   
the evolution approach used  in this work  may 
be  supplemented with nonlinear corrections~\cite{av1,av2,ku1}   
to describe aspects of 
 parton saturation~\cite{satu-fit,kowa-dipole,valpa,costa-nick}. 
It will therefore be  of  interest to  investigate  the extension 
of the   work presented  in this  paper   to low $Q^2$.  

 The   inclusion of  data at higher $Q^2$, 
relaxing the low-$x$ kinematic  cut,   will   constitute  
  a  further  
development  of the TMD formulation.  High-$x$ theoretical 
 issues, including TMD quark distributions,   are 
discussed e.g.~in~\cite{fleming,jain,high-x}.  
Analyses of DIS data over  the whole available 
 range in $x$ and $Q^2$ in terms of TMD pdfs  
will be relevant  for the  calculational  program~\cite{van-ham} 
 of  
off-shell  hard cross sections.

\section{Fits to DIS precision data} 
\label{sec:3} 

The fit to the HERA structure function measurements 
is performed by applying 
the \verb+herafitter+  program~\cite{Aaron:2009aa,herafitter} to 
determine the parameters of the starting distribution ${\cal A}_0$  at the starting scale $q_0$. We perform fits by using two possible 
parameterizations of ${\cal A}_0$: the five-parameter form 
\begin{eqnarray}
x{\cal A}_0(x,\kt) = N x^{-B} \cdot (1 -x)^{C}\left( 1 -D x 
+ E \sqrt{x}   \right) 
   \exp[ - k_t^2 / \sigma^2 ]  \;\; , 
\label{a0-5par}
\end{eqnarray}
and the three-parameter form
\begin{eqnarray}
x{\cal A}_0(x,\kt) = N x^{-B} \cdot (1 -x)^{C}  
 \exp[ - k_t^2 / \sigma^2 ]  \;\; . 
\label{a0-3par}
\end{eqnarray}
As in Eq.~(\ref{gauss}), we take $ \sigma^2  =  q_0^2 / 2 $. 
The parameters $N,B,C,D, E$ (resp.~$N, B, C$) 
in Eq.~(\ref{a0-5par}) 
(resp.~Eq.~(\ref{a0-3par}))
are determined by fitting  the high-precision  structure function 
measurements~\cite{Aaron:2009aa} 
in the range $x<0.005$ and $Q^2>5$~GeV$^2$, and 
the charm production measurements~\cite{comb-charm}, which 
are  in the range 
$Q^2 > 2.5$~GeV$^2$. 
The results presented here are obtained with the \verb+herafitter+ package by 
 treating the correlated systematic uncertainties separately from the uncorrelated statistical and systematic uncertainties.

\begin{table}[ht]
\centering  
\begin{tabular}{|c|c|c|c|}
\hline
       	&$\chi^2 / ndf  ( F_2^{({\rm{charm}})}  ) $	&$ 
\chi^2 / ndf  ( F_2 ) $	& $\chi^2 / ndf  \left( F_2  \mbox{ and } F_2^{({\rm{charm}})}  \right) $	  \\ \hline
3-parameter 		&0.63 	&1.18  &	1.43  \\ \hline
5-parameter		&0.65	&1.16  &	1.41	   \\ \hline
\end{tabular}	
\caption{\it The values of $\chi^2 / ndf $  corresponding to the 
 best fit for charm structure function  $F_2^{({\rm{charm}})} $,   
 for inclusive structure function $F_2$,  and for the 
 combination of $F_2^{({\rm{charm}})}$ and  $F_2$.}
\label{tablechi}
\end{table}

\subsection{Charm  structure function $F_2^{({\rm{charm}})} $ and 
 inclusive structure function $F_2$} 
\label{subsec:3a}

We   fit the  charm leptoproduction 
data~\cite{comb-charm}  and 
the  inclusive structure function 
data~\cite{Aaron:2009aa}   
  based on high-energy factorization and CCFM evolution as 
described in Sec.~\ref{sec:2}. In particular, we also include 
 two-loop running coupling, 
gluon splitting  and consistency  constraint as 
 in Subsec.~\ref{subsec:2a}, and, 
in addition to the  gluon-induced process 
$\gamma^* g^* \to q\bar{q}$,  the contribution from valence quarks is included via $\gamma^* q \to q$ as   in Subsec.~\ref{subsec:2b} 
by using  a CCFM evolution of valence quarks~\cite{Deak:2010gk}.

\begin{figure}[htb]
\begin{center} 
 \includegraphics[scale=0.65]{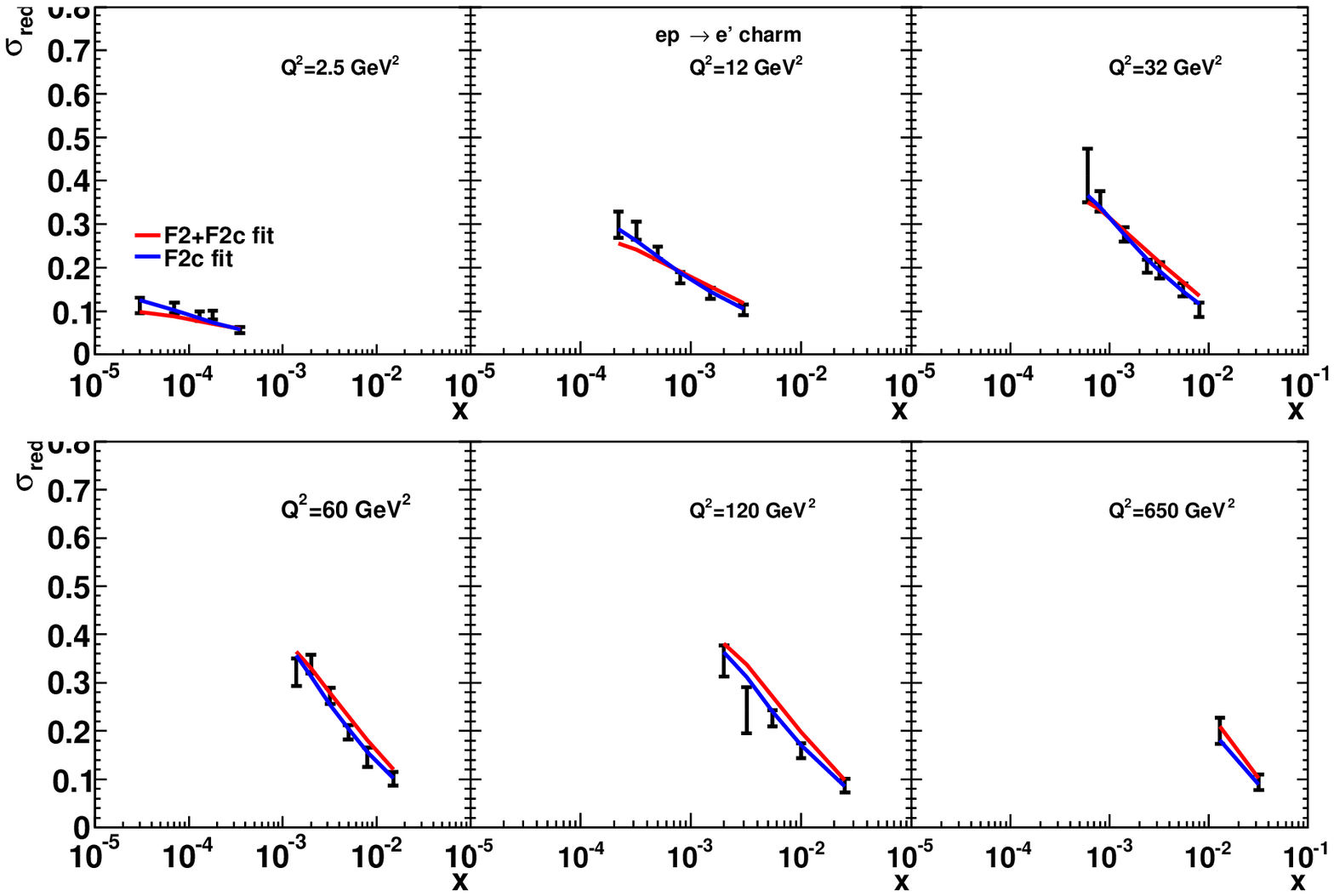}
 \includegraphics[scale=0.65]{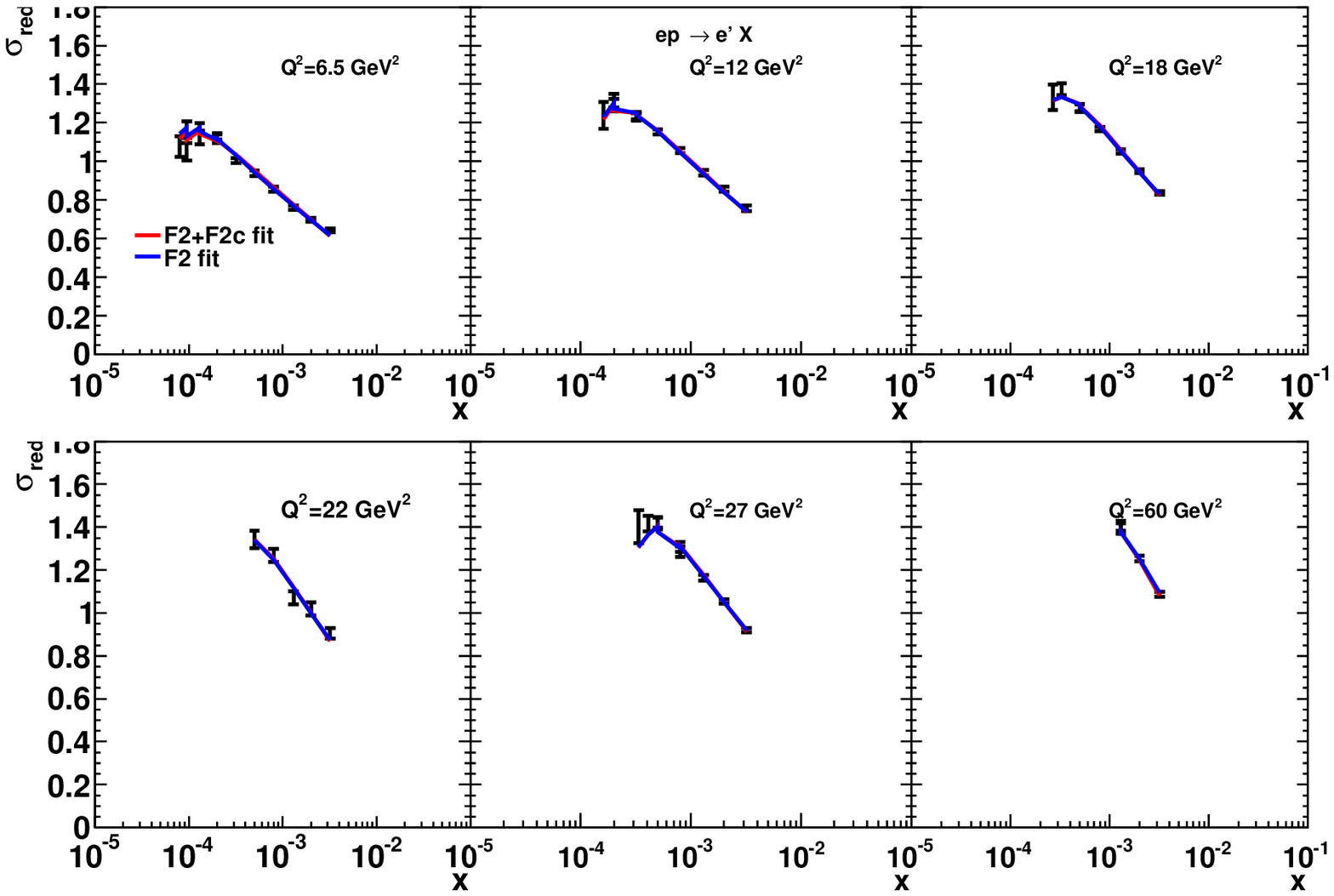}
  \caption{\it The fit to 
DIS high-precision data: 
(top) charm leptoproduction 
data \protect\cite{comb-charm};  
(bottom) inclusive structure function 
data  \protect\cite{Aaron:2009aa}.
}
\label{fig:f2datafit}
\end{center}
\end{figure}

To obtain a reasonable fit to 
structure function data,  we vary the starting scale $q_0$, the QCD scale  $\Lambda_{\rm{QCD}}$ and the charm quark mass $m_c$.   
The  results  in Table~\ref{tablechi} and in 
Fig.~\ref{fig:f2datafit} are obtained for $q_0 = 2.2$ GeV, 
$\Lambda_{\rm{QCD}} = 0.2 $ GeV at $n_f = 4$, $ m_c = 1.45$ GeV. 
Table~\ref{tablechi}  reports the values of $\chi^2$ per 
degree of freedom for the best fit to the charm 
structure function $F_2^{({\rm{charm}})} $~\cite{comb-charm}  
in the full data range 
$Q^2 > 2.5$~GeV$^2$, and 
 to the inclusive structure function 
$F_2$~\cite{Aaron:2009aa} in the data range 
 $x<0.005$,  $Q^2>5$~GeV$^2$ and for a combination of both, in the cases  
of the three-parameter fit  (\ref{a0-3par}) and 
five-parameter fit (\ref{a0-5par}).

The best-fit  $\chi^2 / ndf $  is below 1 for  the charm 
structure function, while it is around 1.18 for the inclusive 
structure function.   This is in accord with the expectation 
 that  charm production is dominated by the gluon distribution, 
while the inclusive structure function receives significant contributions 
from quark channels,  for which an improved treatment at unintegrated 
level is needed.

Also, the gluon density determined from charm measurements 
turns out to  be  steeper at small $x$ than that determined 
from the  inclusive structure function. The resulting 
 tension between the two  fits may be  related to the 
the fact that the starting scale $q_0  $   is 
 not far from the charm threshold  mass $2 m_c $.   
The $\chi^2 / ndf $ for the fit to both $F_2^{({\rm{charm}})} $  
and $F_2$ is 1.43 for the three-parameter fit, and is not 
significantly changed by  using the five-parameter form.

Fig.~\ref{fig:f2datafit} shows the description of the 
charm leptoproduction measurements~\cite{comb-charm} and 
inclusive structure function measurements~\cite{Aaron:2009aa}, 
by the individual  fits to $F_2^{({\rm{charm}})} $ 
and $F_2$ and by the combined  fit.  
Plotted are the reduced cross sections  defined in~\cite{Aaron:2009aa,comb-charm}.

\begin{figure}[htb]
\begin{center} 
 \includegraphics[scale=0.6]{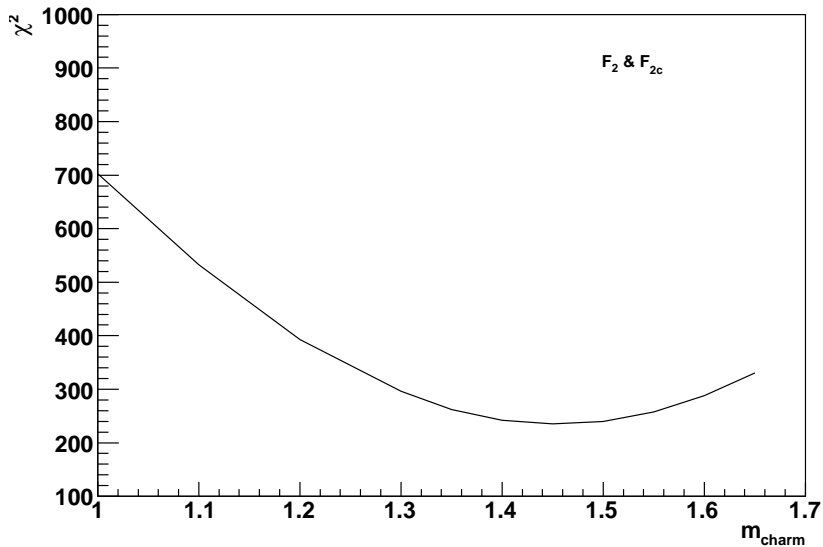}
  \caption{\it The values of  $\chi^2$ for the fit to 
DIS high-precision data including 
charm leptoproduction  \protect\cite{comb-charm} 
versus the charm quark mass. 
}
\label{fig:charmscan}
\end{center}
\end{figure}

In Fig.~\ref{fig:charmscan} we show the results of 
 a scan in the charm mass, by plotting the values of 
$\chi^2$ for the fit to charm and inclusive measurements 
as a function of the charm quark mass. The minimum is 
reached for $m_c = 1.45$~GeV.

\subsection{Unintegrated TMD gluon density} 
\label{subsec:3b}

\begin{figure}[htb]
\begin{center} 
 \includegraphics[scale=0.7]{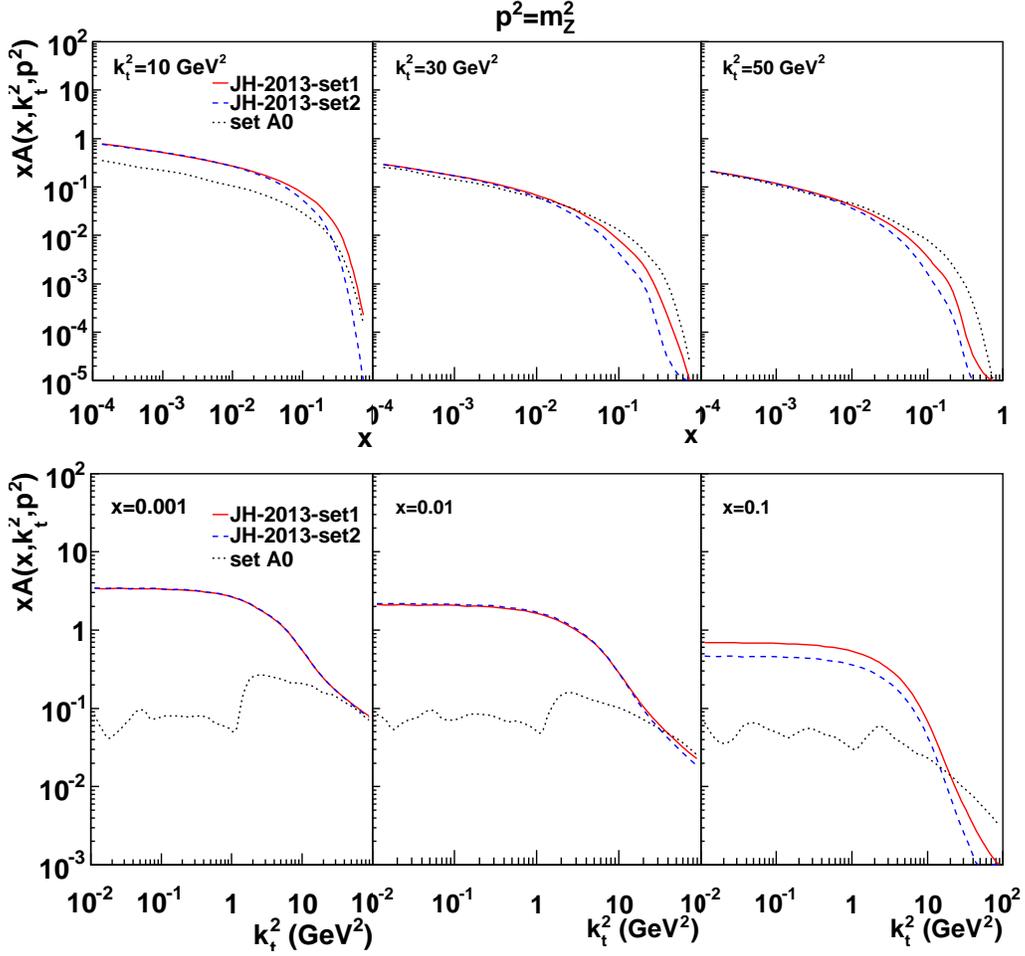}
  \caption{\it  Unintegrated TMD gluon density (JH-2013-set1 
and JH-2013-set2) 
at evolution scale equal to the $Z$-boson mass, $p^2 = m_Z^2$:   
 (top) as a function of $x$ for different values of $\kt^2$;  (bottom)  
  as a function of $\kt^2$ for different values of $x$. 
The results are   compared with 
   {set A0} \protect\cite{Jung:2010si}.
}
\label{Fig:updf}
\end{center}
\end{figure}

\begin{figure}[htb]
\begin{center} 
 \includegraphics[scale=0.7]{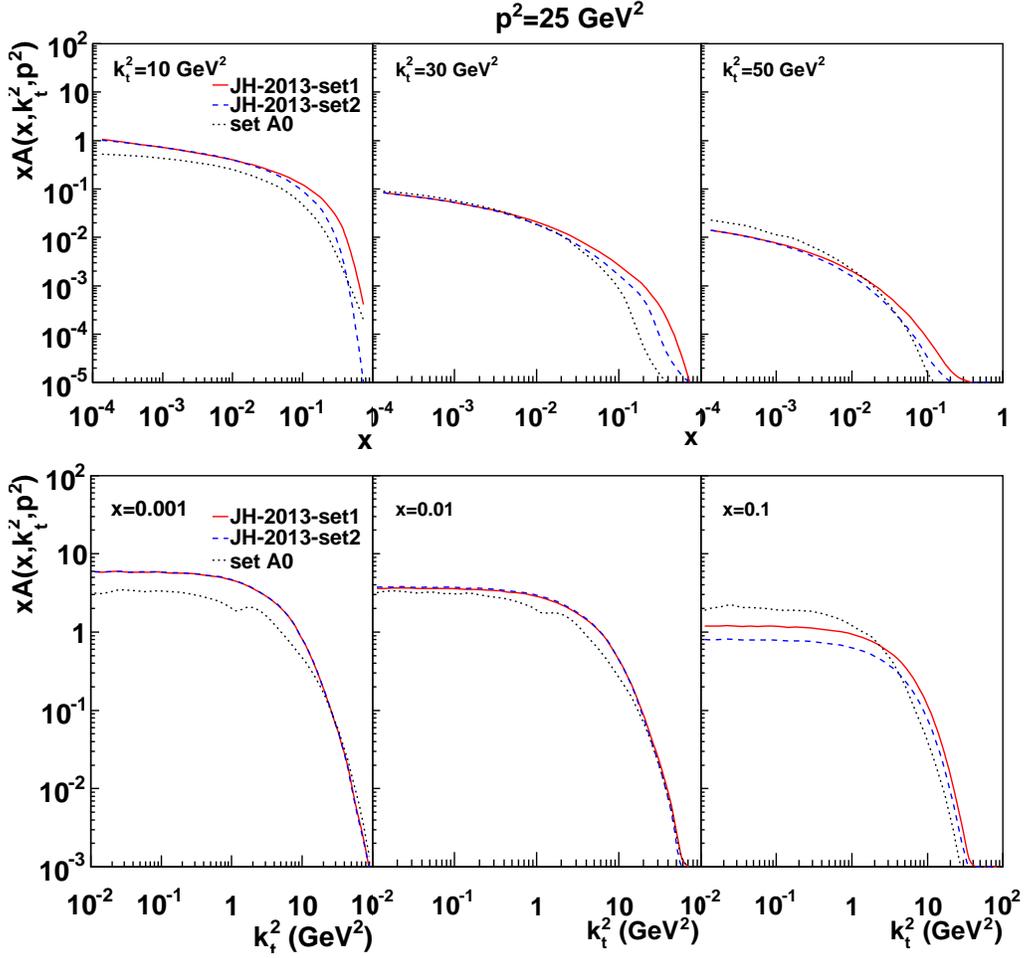}
  \caption{\it  Unintegrated TMD gluon density (JH-2013-set1 
and JH-2013-set2) 
at evolution scale  $p^2 = 25$ GeV$^2$:   
 (top) as a function of $x$ for different values of $\kt^2$;  (bottom)  
  as a function of $\kt^2$ for different values of $x$. 
The results are   compared with 
   {set A0} \protect\cite{Jung:2010si}.
}
\label{Fig:updfbis}
\end{center}
\end{figure}

 Here we 
 present two sets of unintegrated  pdfs 
determined   from the 
fits to  high-precision  DIS 
measurements described  in the previous subsection:  
JH-2013-set1 is  determined
from the fit to  inclusive 
$F_2$  data  only; 
 JH-2013-set2 is  determined 
from the fit to both $F_2^{({\rm{charm}})} $  
and $F_2$ data. 
 The unintegrated TMD gluon density is shown in 
 Figs.~\ref{Fig:updf}   and   \ref{Fig:updfbis}  
at  different evolution scales, versus the longitudinal 
momentum fraction $x$ and versus the transverse momentum 
$\kt$. The results  are compared with   the  older 
parton distribution set A0~\cite{Jung:2010si}, which    
did  not use  the 
precision data and 
did  not include   two-loop running coupling, 
kinematic consistency constraint, 
nonsingular terms in the gluon 
splitting function,  
 and  unintegrated valence 
 quark density.

We see from the evolution      to   scale 
$p^2$ equal to the $Z$-boson mass  in Fig.~\ref{Fig:updf}
that the   fit to the DIS high-precision measurements 
gives significant differences in the TMD gluon  density 
  compared to 
earlier determinations, especially in the region of medium to 
low  $\kt$.  For the 
 lower  $p^2$ scale in  Fig.~\ref{Fig:updfbis} the differences are 
less pronounced but still important especially 
for small values of $x$.

\subsection{Uncertainties on the TMD gluon density} 
\label{subsec:3c}

In this subsection we  consider 
 separately experimental and theoretical uncertainties of 
the  TMD parton distributions.

\begin{figure}[htb]
\begin{center} 
 \includegraphics[scale=0.38]{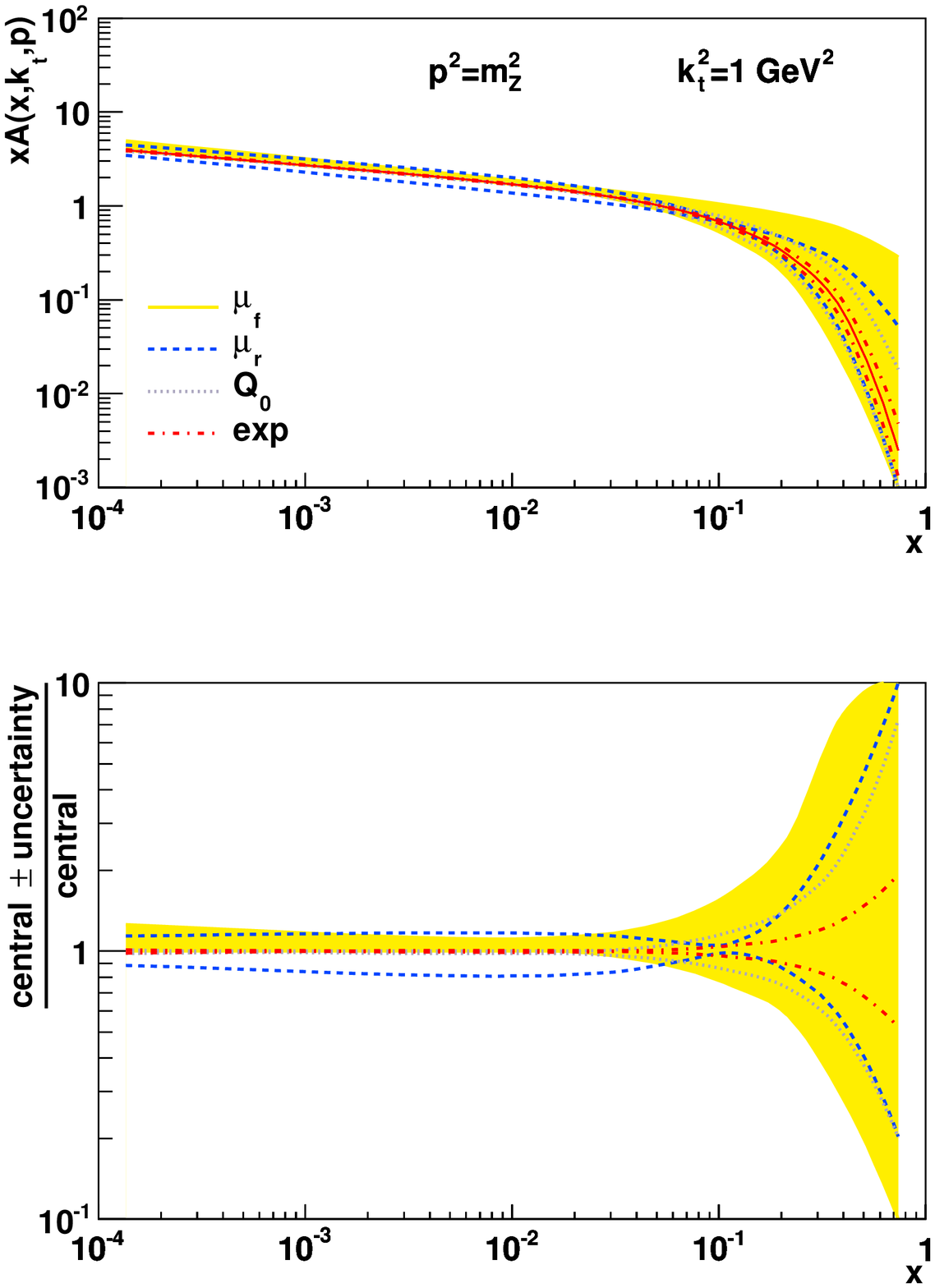}
 \includegraphics[scale=0.38]{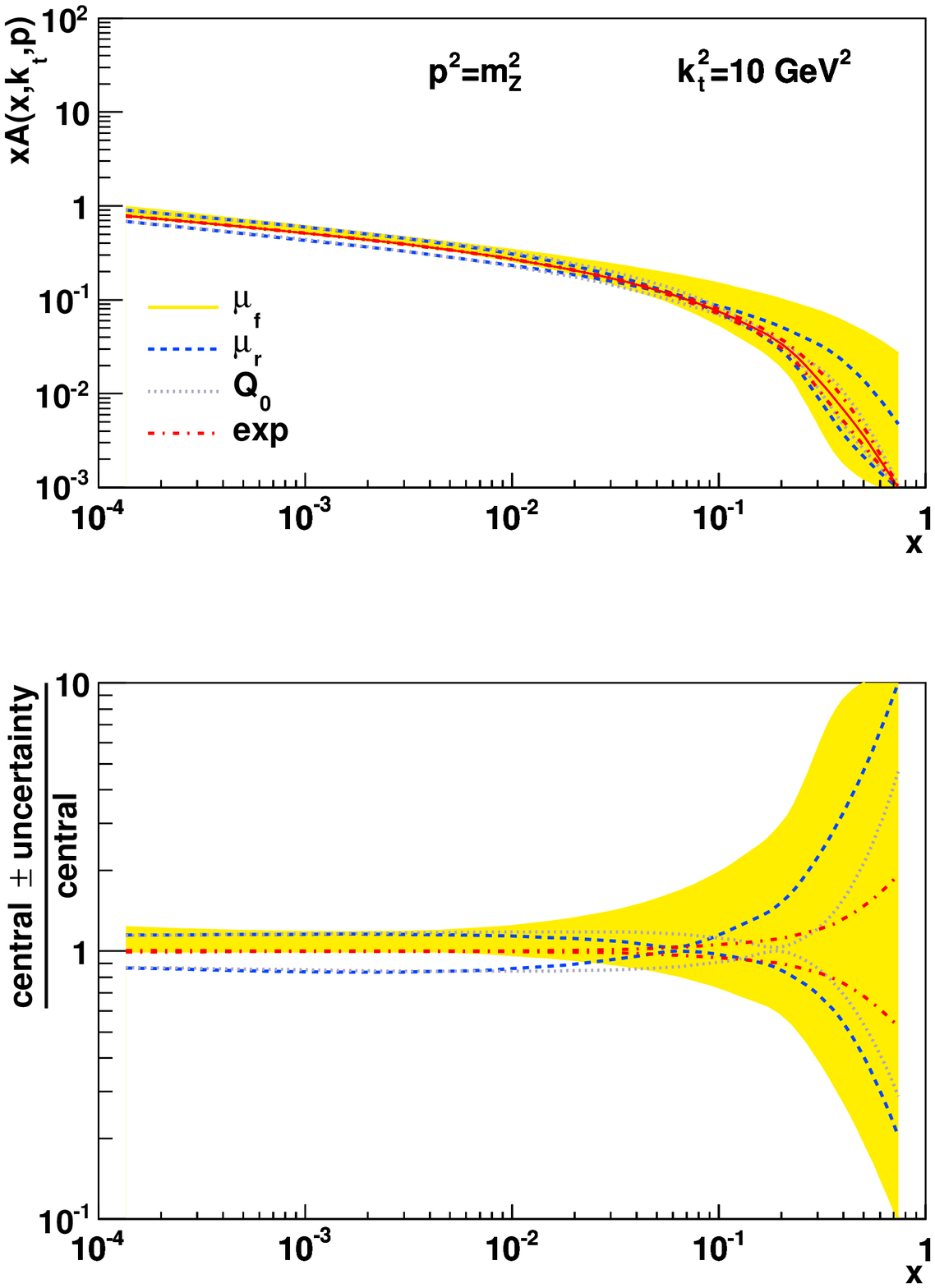}
  \caption{\it Experimental and theoretical 
uncertainties of the unintegrated TMD gluon density versus $x$ 
for different values of transverse momentum 
at  $p^2 = m_Z^2$. The yellow band gives the uncertainty 
from the factorization scale variation; the  curves indicate  
the uncertainties from the other sources. 
}
\label{Fig:updf-uncertainty}
\end{center}
\end{figure}

\begin{figure}[htb]
\begin{center} 
 \includegraphics[scale=0.38]{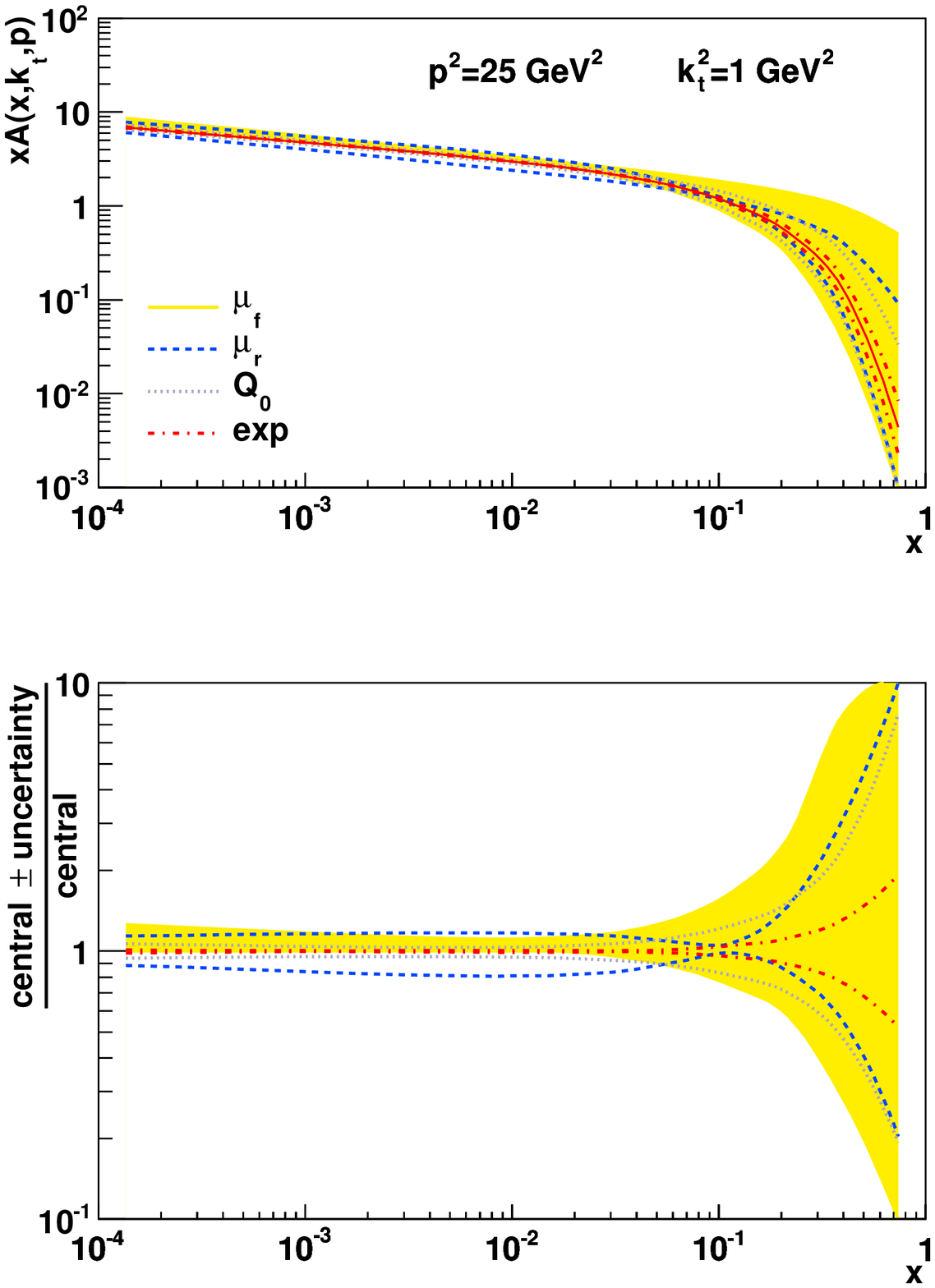}
 \includegraphics[scale=0.38]{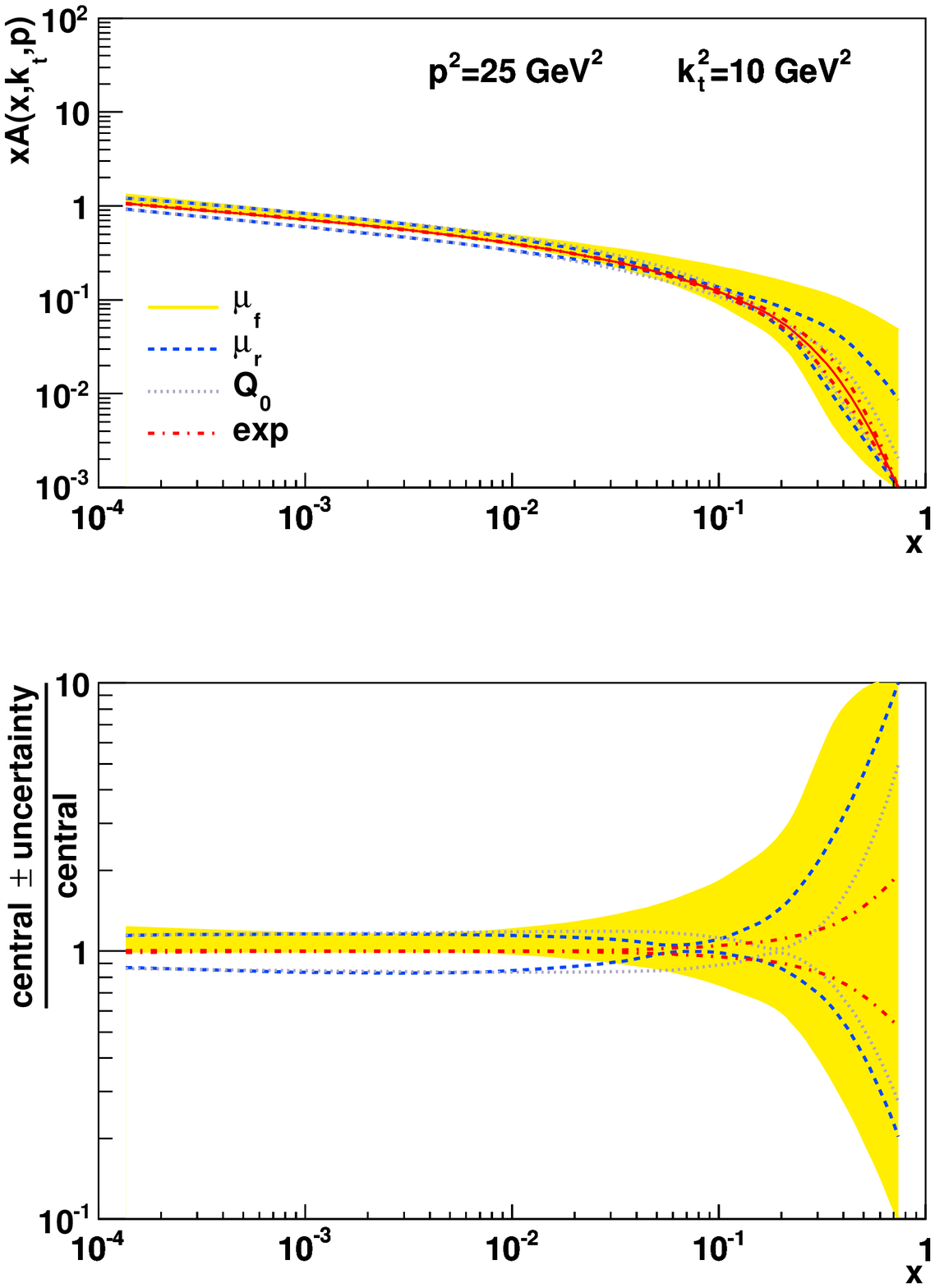}
  \caption{\it Experimental and theoretical 
uncertainties of the unintegrated TMD gluon density versus $x$ 
for different values of transverse momentum 
at  $p^2 = 25$ GeV$^2$. The yellow band gives the uncertainty 
from the factorization scale variation; the  curves indicate  
the uncertainties from the other sources. 
}
\label{Fig:updf-uncertainty-bis}
\end{center}
\end{figure}

Experimental uncertainties are obtained 
  within the  \verb+herafitter+ package from a variation of the individual parameter uncertainties,  following the procedure 
  described in~\cite{Pumplin:2001ct} applying $\Delta \chi^2=1$. 
These result in 10 to  20 percent gluon uncertainty for medium and large $x$. 
The experimental  uncertainties on the gluon at small $x$ 
are    small (much smaller than those obtained in standard   fits based on integrated pdfs),  
since only the gluon density is fitted. The  
experimental uncertainties  are shown 
by the dot-dashed red curves 
in Fig.~\ref{Fig:updf-uncertainty} as a function of $x$ 
for different values of transverse momentum at the
 evolution scale $p^2$ equal to the $Z$-boson mass.
Similarly in Fig.~\ref{Fig:updf-uncertainty-bis} for lower 
evolution scale $p^2 = 25$ GeV$^2$. 

Next we consider 
 theoretical uncertainties.  The first  such kind of  uncertainty is the 
dependence  on  the starting scale $q_0$ for  gluon density evolution.  
In  Figs.~\ref{Fig:updf-uncertainty}  
and~\ref{Fig:updf-uncertainty-bis}   
 the dotted blue curves 
 show  the  effect  
on the  gluon distribution  from  variation in the 
starting scale  $ q_0 $. These  uncertainties 
are small at small $x$, while they become very 
large at large $x$ because in this region,  
since we fit    $F_2$ 
in the range $x<0.005$ and $Q^2>5$~GeV$^2$,   there is 
little   constraint from data.

We also consider theoretical  uncertainties on the 
TMD gluon density  from  variation of 
the factorization scale  and renormalization scale.  
This  approach is  different   from that usually 
 followed in 
determinations of ordinary, collinear 
 pdfs from   fixed-order perturbative 
treatments~\cite{pdf4lhc-alekhin}. 
In this case, no uncertainty on the pdfs is considered from 
scale variation. Only  when computing predictions for 
any specific observable 
 the theoretical uncertainty on  the 
predictions  is estimated by   scale variation.  In our 
approach we are interested to estimate the uncertainty 
from varying scales in the theoretical calculation  used  
to determine the pdf.  In 
Figs.~\ref{Fig:updf-uncertainty}    
and~\ref{Fig:updf-uncertainty-bis} 
  the 
renormalization scale (blue dashed curves) and 
 in the factorization 
scale  (yellow band)  are varied by  a factor of 2.

\subsection{Integrated parton distributions} 
\label{subsec:3d}

\begin{figure}[htb]
\begin{center} 
 \includegraphics[scale=0.38]{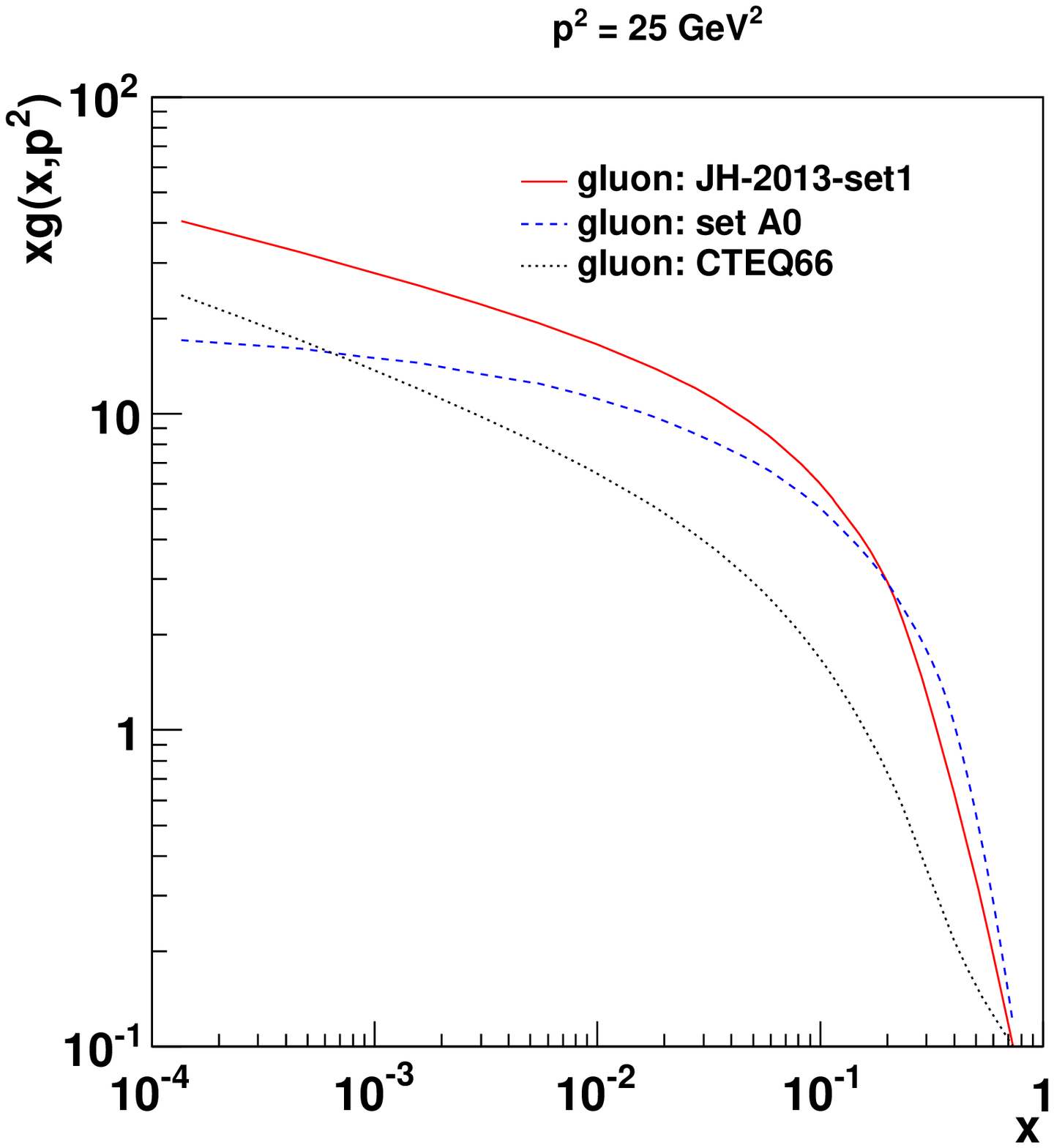}
 \includegraphics[scale=0.38]{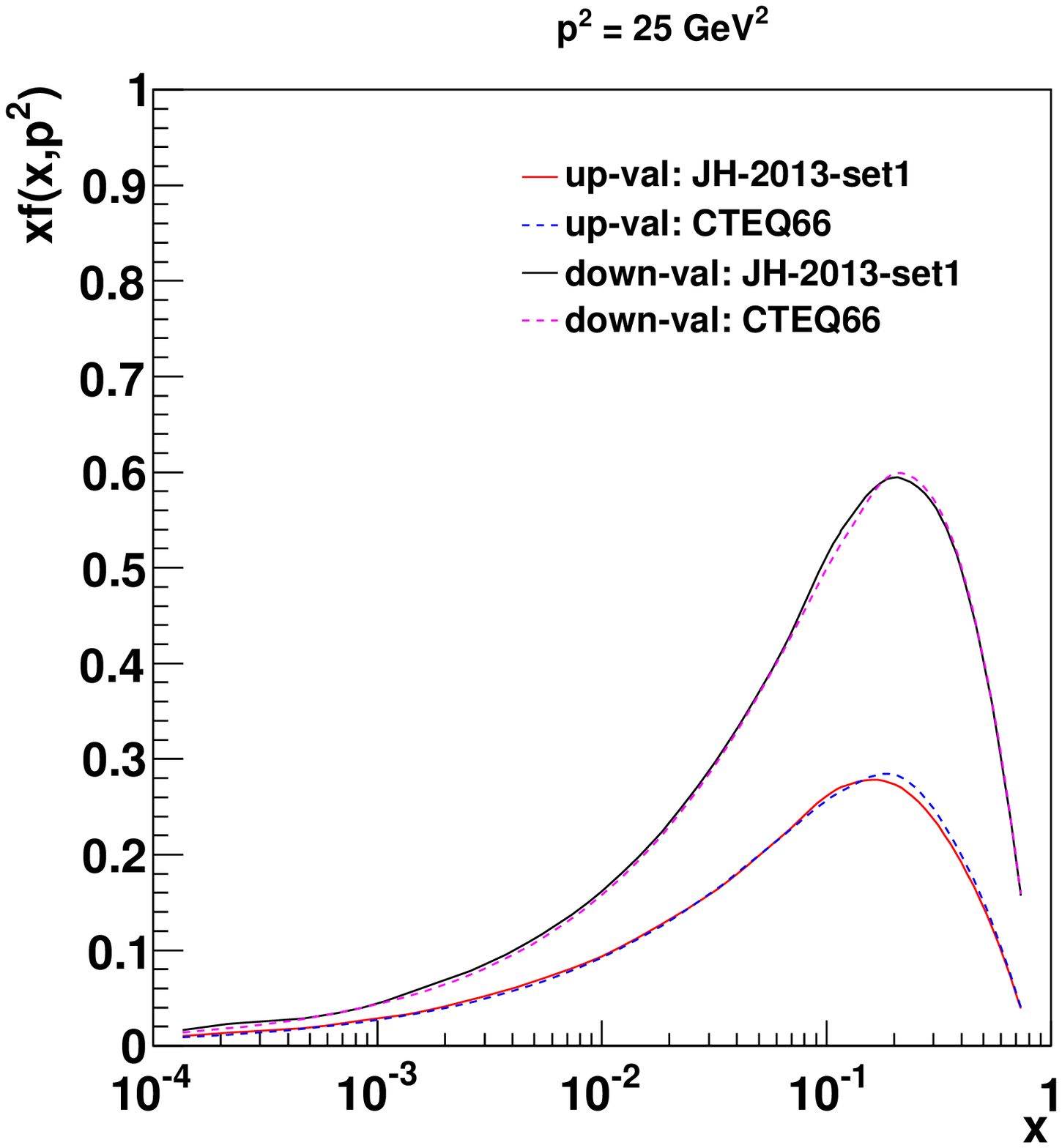}
 \includegraphics[scale=0.38]{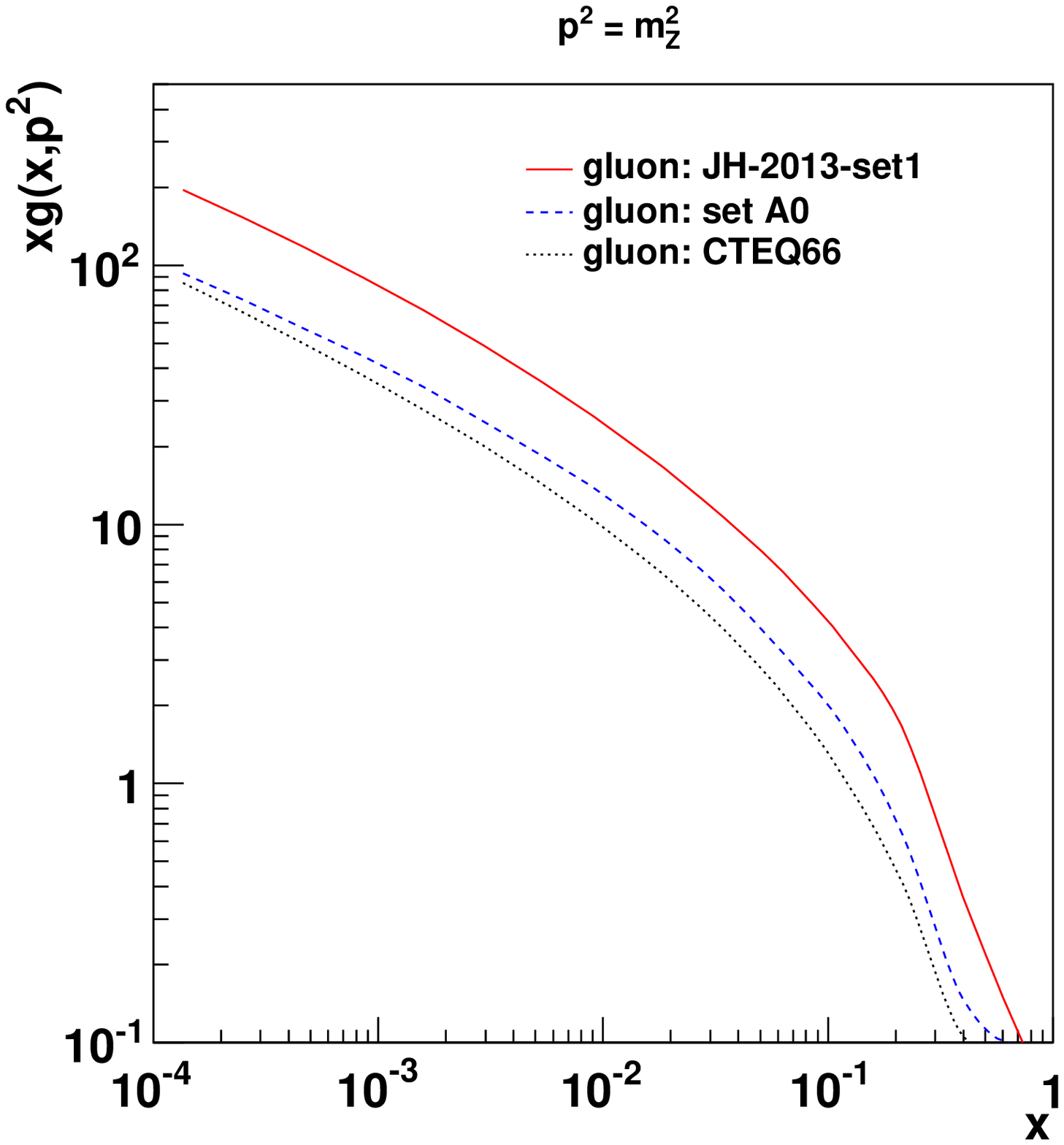}
 \includegraphics[scale=0.38]{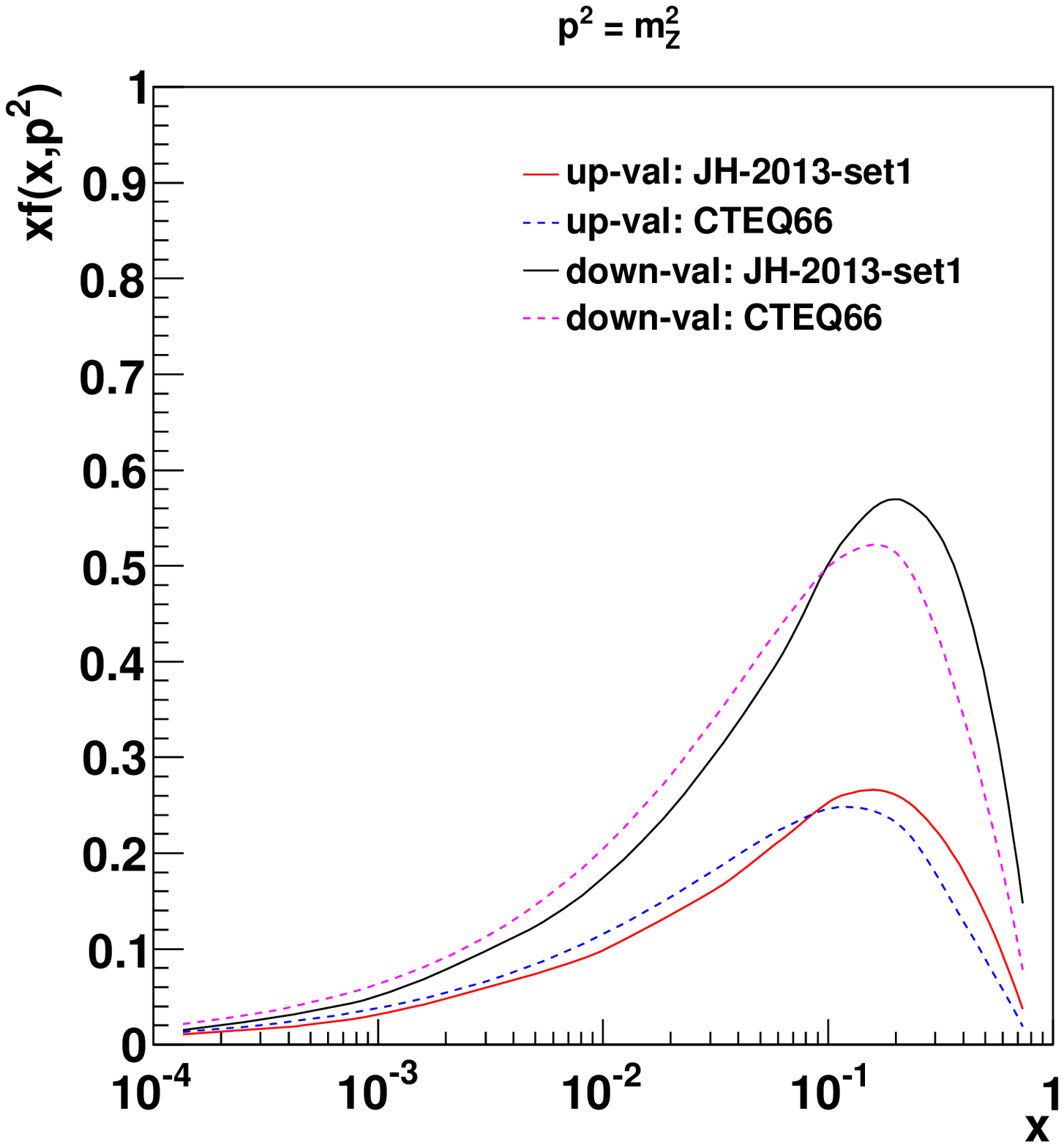}
  \caption{\it Integral over transverse momenta of the TMD distributions  
for (left) gluon and (right) valence  quark  
at  different evolution scales: (top) $ p^2 = 25 $ GeV$^2$; 
(bottom) $ p^2 = m_Z^2 $. 
}
\label{fig:integrated}
\end{center}
\end{figure}

For a  cross check with the integrated pdfs we  now compute the integral over transverse 
momenta of the TMD parton distributions. In Fig.~\ref{fig:integrated} 
we plot the results for  gluon and    valence quark distributions, 
obtained 
from the set JH-2013-set1 of this paper (and also,  for comparison, from  the gluon 
 in the older set A0~\cite{Jung:2010si}), at two  
different evolution scales. For comparison we   
plot  the ordinary,  integrated distributions 
obtained from the NLO-DGLAP CTEQ 6.6~\cite{nado08} fit.  
We observe good agreement for the integral of  
valence quark distributions at low scales while differences arise from the different evolution at larger scales. For 
  the gluon case, the  difference between 
 the integral of TMD and CTEQ reflects the shuffling 
 of  flavor singlet  contributions  between  sea quark and gluon  
 in the two formalisms.

We conclude this section by noting that the 
CCFM evolution  kernel can be approximated to one loop 
by using collinear ordering~\cite{hj04,hj-ang}.  This constitutes the  
DGLAP  limit of the evolution equation. 
If we  perform fits to the high-precision $F_2$ data by using 
 Eqs.~(\ref{kt-factorisation}),(\ref{uglurepr1}) in the one-loop approximation mode  
  we find that this   approximation   is unable  to give a good fit  based on the TMD gluon only, 
   $\chi^2/ndf \sim 6$. 
We interpret this as a check on the consistency of the physical 
picture,  signaling 
 the need for  introducing quark-initiated  processes in the 
collinear framework.

\section{TMD gluon density at the LHC}
\label{sec:4}

The   TMD  parton distributions  determined in  Sec.~\ref{sec:3} 
from fits to the high-precision DIS data can be used to make predictions for 
 hadron-hadron collider processes. 

An example is the  Drell-Yan  vector boson production. 
We here consider  $ W  $-boson production in association with jets. 
This process   is important both    for 
standard model   physics and for new physics searches at the LHC.  
In particular  it  is relevant to    studies of 
parton distribution functions and 
 of Monte Carlo event generators~\cite{pdf4lhc-alekhin}, 
including   signals of   multi-parton interactions,  for which  
$ W + 2$ jets is a classic  channel~\cite{ellie,paolo}.

\begin{figure}[htb]
\begin{center} 
 \includegraphics[scale=0.38]{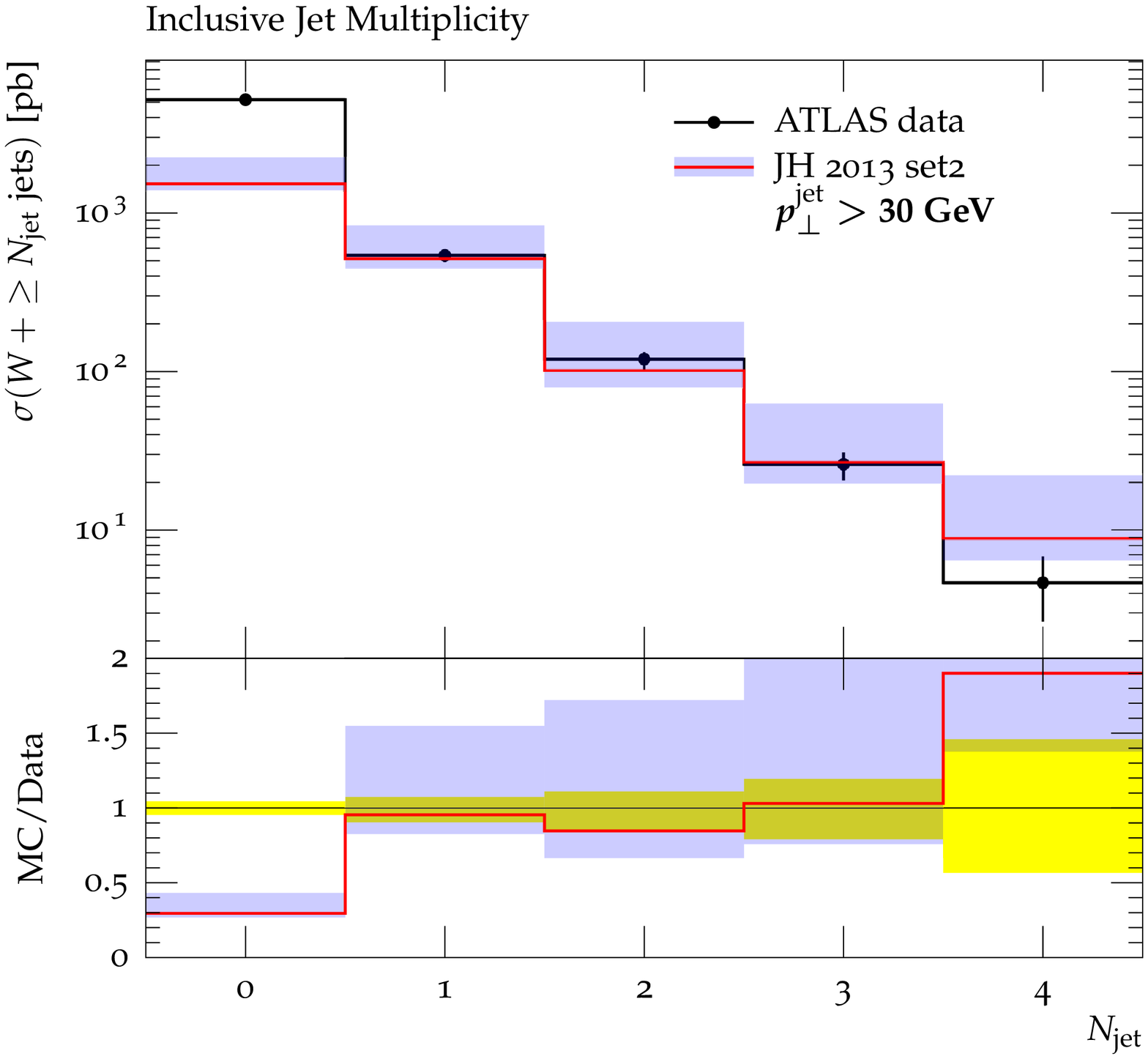}
 \includegraphics[scale=0.38]{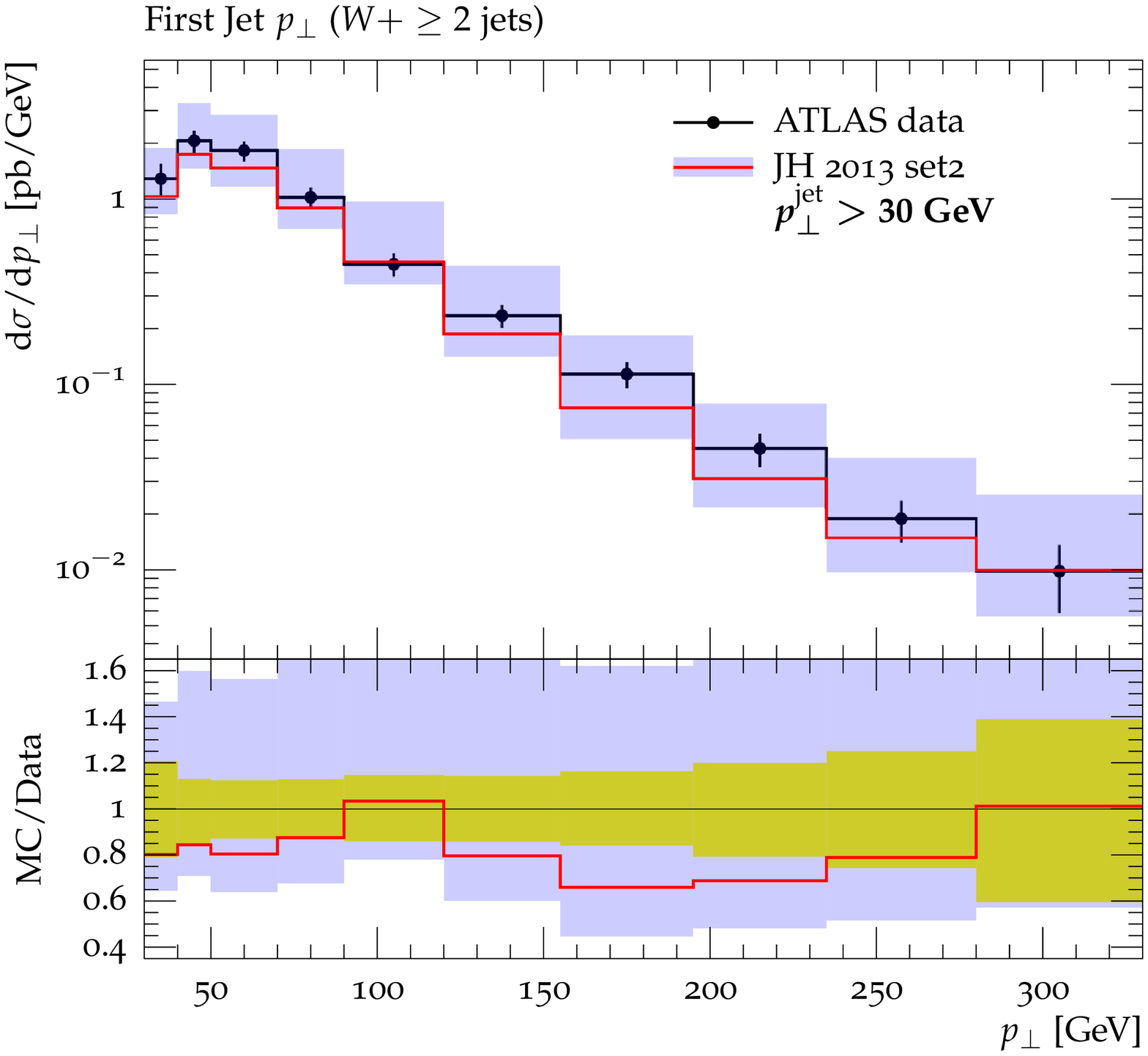}
  \caption{\it Predictions for $W + $ jets production using the  
unintegrated TMD gluon density JH-2013-set2: 
(left) inclusive jet multiplicity; (right) leading jet 
transverse momentum spectrum. The experimental data are 
 from \protect\cite{atlas-w-jets}. The yellow band is 
the experimental  uncertainty. 
The blue band is the theory uncertainty.}
\label{fig:wjets}
\end{center}
\end{figure} 

 To compute predictions  for     $ W  $-boson + jets final  states~\cite{w-j-prepar}, we 
 use the CCFM gluon and valence quark 
distributions determined in the previous section,    convoluted with  
high-energy matrix 
elements~\cite{hent12,marball} with off-shell 
partons~\cite{fad06,lip01} for weak boson production.  We 
  present  results for the inclusive jet multiplicity 
distribution and   leading  jet transverse momentum spectrum.   
 The results (obtained with the {\sc Rivet} - package~\cite{Buckley:2010ar}) 
 are  shown in Fig.~\ref{fig:wjets} along with the ATLAS 
 measurements~\cite{atlas-w-jets}.  

The solid red curves  in 
Fig.~\ref{fig:wjets} 
  are the predictions from  JH-2013-set2, with the blue band    
  corresponding to the  pdf uncertainty. 
Both the jet  multiplicity and the 
 transverse momentum are reasonably well  
 described by    the  predictions.   Further discussion on $W$ + jets 
will be given in~\cite{w-j-prepar}.   The production of   final states with 
$W$ boson and multiple  jets  at the LHC  receives sizeable 
contributions  from 
 large  separations in rapidity between final-state particles. 
However, the cross  sections  computed in Fig.~\ref{fig:wjets} 
 are not 
dominated by very small values of $x$. As a result,  
  the uncertainty band due to the 
uncertainties in the TMD pdfs  is significant. 
A comparison with NLO-matched results and 
corresponding uncertainties  is presented in~\cite{w-j-prepar}. 
 It is conceivable that 
combining $pp$ measurements on vector boson production 
with the DIS measurements  analyzed in this paper  may help  
to constrain   TMD  pdfs  especially at medium to large values of $x$.

\section{Conclusions} 
\label{sec:5}

In this work  we have performed the  first determination of  the
TMD gluon density function   from   high-precision  DIS measurements,  
 including  experimental and theoretical    uncertainties.

We have presented fits,   
based on    
QCD high-energy factorization and CCFM evolution, 
 to   HERA  charm-quark  leptoproduction data for  the  
  structure function 
$F_2^{({\rm{charm}})} $~\cite{comb-charm} in the range 
$Q^2 > 2.5$~GeV$^2$, 
and to HERA 
$F_2$ structure function 
data~\cite{Aaron:2009aa} in the  range 
$x<0.005$ and $Q^2>5$~GeV$^2$. 
In this  approach  the charm   structure function 
can be regarded as a physical probe of the unintegrated 
 TMD  gluon density.   We fit the combined HERA 
charm-quark 
data~\cite{comb-charm} over the whole kinematic range of the 
measurement, and obtain 
that 
the best fit gives  $\chi^2$ per degree of freedom 
       $\chi^2/ndf \simeq 0.63$. 
The inclusive  $F_2$  structure function 
involves both gluon-density and quark-density channels. 
We fit the combined HERA 
$F_2$  
data~\cite{Aaron:2009aa} in the   kinematic  range 
$x<0.005$, $Q^2>5$~GeV$^2$, and  obtain 
that the best fit  gives         $\chi^2/ndf \simeq 1.18$. 
Despite the restricted kinematic range, 
the great  precision  of the data provides a 
highly  nontrivial test  of the approach.  
We find  a good     fit 
to  both charm-quark and inclusive 
data.  
Based on this,  we make a determination of the TMD gluon 
density  (as well as of the QCD scale $\Lambda_{\rm{QCD}}$ 
 and the charm mass $m_c$)  and  present   new 
unintegrated pdf sets,  JH-2013.  
As a result of  the high-precision data, 
the   JH-2013  distributions  differ significantly from earlier sets. 
We also present   experimental and theoretical 
uncertainties associated with the  
TMD  pdfs.   We  compute  predictions based on the  TMD pdfs  
  for $W$-boson plus jets production at the 
LHC, and find that the  results compare well with the 
measurements~\cite{atlas-w-jets} of 
jet multiplicities and transverse momentum spectra within the pdf uncertainties.

The  approach  of this work 
is based on the use of transverse momentum 
dependent matrix elements and evolution. 
Both the kernel and the initial condition 
of  the    evolution equation   are  $k_t$-dependent. 
The transverse momentum dependence of the gluon 
density  arises from  both   
perturbative and  nonperturbative processes. 
 The physical picture of DIS scaling violation 
underlying this  approach differs from that  
of finite-order perturbative  QCD fits, e.g. at the NLO level, because it 
 takes into account  corrections 
to the collinear ordering in the initial state evolution 
to all orders in the QCD coupling 
$\alpha_s$.  On the other hand,  it 
 also differs from BFKL evolution 
because it takes into account, for any $x$,  color  coherence 
associated with   soft  multi-gluon emission.  
 In this paper   we   have 
developed   a parton branching  Monte Carlo  implementation  of  
 the CCFM  evolution 
equation  and  
 we have  included it   in the 
 \verb+herafitter+  
 program~\cite{Aaron:2009aa,herafitter}.

The   choice  of the kinematic range for  the $F_2$  data 
considered in this paper  is   motivated 
by  the fact that our approach relies  on  
 perturbative factorization theorems, which 
classify higher-order  corrections 
 according to  logarithmic hierarchy  
based on  high  $Q^2$ and low  $x$. 
However,  we   note    that  
the choice made in this paper  is   conservative,  
 and   the  physical picture   
lends itself to extensions to  lower $Q^2$ and higher $x$. 
On one hand, 
this   picture  goes beyond DGLAP   by including 
non-collinear emissions to all orders in $\alpha_s$.  
On the other hand, 
 it goes beyond BFKL   by including  large-$x$ terms 
according to  the  CCFM prescription. 
Furthermore, the study of  $W$-boson + jets made in this paper 
suggests that   $pp$  measurements  of 
 vector boson production at the LHC 
may be used   to extend  experimental investigations  of TMD 
 parton density functions.

\section*{Acknowledgements}

Many  thanks to the HERAfitter team for  
 invaluable  advice.  
We acknowledge useful discussions and comments on the 
paper draft from  M.~Cooper-Sarkar, 
  S.~Glazov,  R.~Placakyte, S.~Prestel  and V.~Radescu.  
FH thanks the Terascale Physics Helmholtz Alliance and DESY 
for  support and hospitality.

\end{document}